\documentclass[5p,final]{elsarticle}
\usepackage{amsfonts}
\usepackage{amssymb}
\usepackage{amsmath}  
\usepackage{amsthm}
\usepackage{enumitem} 
\usepackage{geometry}
\usepackage{graphicx}
\usepackage{algorithm}
\usepackage{algcompatible}
\usepackage{algpseudocode}
\usepackage[colorlinks=true, allcolors=blue]{hyperref}
\usepackage[utf8]{inputenc} 
\usepackage[english]{babel} 
\usepackage{url}
\usepackage{wrapfig}
\usepackage{subfiles}
\usepackage{xfrac}
\usepackage{multirow}
\usepackage{cleveref}
\usepackage{balance}
\usepackage{svg}
\usepackage{booktabs}
\usepackage{tikz}
\usetikzlibrary{shapes.geometric, arrows}
\usepackage{tikz-3dplot}
\usepackage[squaren]{SIunits}
\usepackage{caption}
\usepackage{multicol}
\usepackage{makecell}
\usepackage{rotating}
\usepackage{tabularx}
\usepackage[normalem]{ulem}
\usepackage{subcaption}
\usepackage{lipsum}
\usepackage{todonotes}
\usepackage{float}
\usepackage{mathtools}

\newcommand{\bc}{\begin{center}}
\newcommand{\ec}{\end{center}}
\newcommand{\be}{\begin{equation}}
\newcommand{\ee}{\end{equation}}
\newcommand{\bea}{\begin{eqnarray}}
\newcommand{\eea}{\end{eqnarray}}
\newcommand{\beq}{\begin{eqnarray*}}
\newcommand{\eeq}{\end{eqnarray*}}
\newcommand{\bv}{\left( \begin{array}{c} }
\newcommand{\ev}{\end{array} \right) }

\usepackage{natbib}
\begin{document}
\title{Metaorder modelling and identification from public data}
\author[uct-sta]{Ezra Goliath}
\ead{gltezr001@myuct.ac.za}
\author[uct-sta]{Tim Gebbie}
\ead{tim.gebbie@uct.ac.za}
\address[uct-sta]{Department of Statistical Science, University of Cape Town, Rondebosch 7701, South Africa}

\begin{abstract}
Market-order flow in financial markets exhibits long-range correlations. This is a widely known stylised fact of financial markets. A popular hypothesis for this stylised fact comes from the Lillo-Mike-Farmer (LMF) order-splitting theory. However, quantitative tests of this theory have historically relied on proprietary datasets with trader identifiers, limiting reproducibility and cross-market validation. We investigate whether it can be recovered from anonymous public data using synthetic metaorder reconstruction. Using transaction and quote data for the largest 239 stocks by market capitalisation on the JSE as of 13 March 2026 with the data range being 1 January 2023 until 31 December 2025, we conduct a grid search over reconstruction parameters and evaluate each configuration against established metaorder stylised facts and the LMF relation. 
Configurations selected to minimise errors across the metaorder impact stylised facts reproduce the targeted aggregate properties but yield a poor LMF relation. Configurations selected to minimise the LMF discrepancy recover the relation by construction while retaining several broad impact features, although some stock-level execution and decay fits are weaker. These asymmetric results show that recovering aggregate impact stylised facts alone is insufficient to identify LMF-consistent order splitting, while the LMF-targeted result establishes compatibility within the reconstruction class rather than an independent test. The findings support consistency with, rather than direct validation of, the LMF theory using anonymous market data.

\end{abstract}


\maketitle
\tableofcontents

\section{Introduction} \label{sec:intro}

Long-range autocorrelations (LRC) of market-order fl\-ow is a widely known macroscopic phenomenon of financial markets \cite{LilloFarmer2004,LilloMikeFarmer2005}. The idea is that macroscopic phenomena in continuously traded double-auction intraday financial markets are predominantly due to the aggregated effects of microscopic actions {\it i.e.} the bottom-up aggregation of individual trades. This can be contrasted both with top-down causes such as market wide information shocks, or the coordinated actions of agents at the macro-economic level \cite{WilcoxGebbie2014} and prices that result for Walrusian like closing auctions. 

One hypothesis is that LRC arise in the market-order flow due to the order-splitting behaviour of traders \citep{LilloMikeFarmer2005}. The key microscopic contributor is then the aggregation of metaorders after they have been broken up and mixed in real markets. Even if the reason (or the cause) for the metaorders themselves is top-down from market actors operator at lower frequency at the level of mutual funds and macro-economic events. 
This top-down cause can be represented through latent supply and demand in the order book \citep{Mastromatteo2014AgentBasedLiquidity,Donier2015FullyConsistentImpact}. 

This idea then links the microscopic properties of tr\-ades directly to macroscopic averaged measurables; the trade-sign auto-correlations and the shape of the price impact. Such a direct link should be surprising if one places more value on exogenous information shocks, fundamental information, and general news flow, and then some sort of idea of predictability in stock market price changes themselves, rather then endogenous causes of dynamics and emergence. 

Concretely, it was proposed that the decay exponent $\gamma$ of the autocorrelation function (ACF) of the trade signs (say $\epsilon$) are directly related to the power-law exponent $\alpha$ of the probability of finding a metaorder of length $L$ through the relation: $\gamma \approx \alpha - 1$ \citep{LilloMikeFarmer2005}. 

This result is known as the LMF theory \footnote{That the probability of metaorders of length $L$ is $P(L) \propto L^{-\alpha-1}$ for $\alpha>1$, and the autocorrelations in the trade signs $\epsilon$ are $C(\tau) \propto \tau^{-\gamma}$ for a delay $\tau$ such that $\gamma=\alpha-1$.} and was initially argued via simulation, but was later empirically validated on the Tokyo Stock Exchange (TSE) \citep{SatoKanazawa2023LMFTest,SatoKanazawa2023}. 

This validation is a remarkable and surprising result! The importance of this result has been discussed elsewhere \cite{Lillo2023Decoding,Bouchaud2025UniversalLaw}, but essentially it is a phenomenological law that explains the concavity of the price impact and hence the square-root law of the price impact of metaorders. It also supports a very specific microscopic cause for the key macroscopic universal features of intraday stock market dynamics and how the order-flow can be predictable even when the price changes are not. 

Verification of the LMF theory on other markets following \citeauthor{SatoKanazawa2023LMFTest} has proven to be difficult because of the need to acquire datasets that contain trader identification information and broker codes. These fields are required to be able to reconstruct the underlying metaorders from the market transaction and limit order update messages. Here a metaorder is a linked sequence of buy (or sell) orders $+1 (-1)$ coming from a single agent operating in the market. There are then many such agents and hence many metaorders. We label the metaorders with index $j$ to refer to the $j$-th metaorder $M^{(j)}$ comprising of $n_j$ orders $m^{(j)}_i$ for $i\in[1:n_j]$. A metaorder could be considered to represent the simplest type of trading schedule. The details of the parent order generating the metaorders are typically considered to be private and hence not made more generally available to researchers and market participants.

However, recent work on metaorder reconstruction using very simple distributional assumptions has proved to have been able to preserve key aggregate properties of financial markets \cite{Maitrieretal2025}. Here we combined these two innovations to address the data issue and reverse engineer the individual trades to generate synthetic metaorders from publicly available Johannesburg Stock Exchange (JSE) data for the largest 239 stocks (Table \ref{tab:239 tickers}). This was carried out over a three year period, from 1 January 2023 to 31 December 2025 using trade by trade Level 1 event data (See Appendix \ref{app:data-metadata}). The data includes only publicly available trades and order-book updates, and was sourced through the BMLL Data Lab platform \cite{bmllpython2026}.  

In Section \ref{sec:data} we provide details about the data source and how the data was pre-processed. In Section \ref{sec:synthetic} we describe how we have implemented the method of \citeauthor{Maitrieretal2025}, and then provide some of the missing algorithm details \cite{goliath_github_repo}. We explicitly provide two simple variants of the metaorder reconstruction method. Part of the metaorder reconstruction method requires choosing the number of effective traders operating within the market and the other part is making distributional assumptions about the meta\-orders themselves. In Section \ref{sec:calibration} we provide details on how we performed a grid search to calibrate the metaorder generating algorithms in Section \ref{sec:synthetic}. In Section \ref{sec:lmf theory} we describe the methods we used to estimate $\alpha$ and $\gamma$, following the broad idea of \citeauthor{SatoKanazawa2023LMFTest} but with the synthetic metaorders following our implementation of the method of \citet{Maitrieretal2025}. Section \ref{sec:calibration results} summarises and provides some insights the results from the grid search and subsequent calibration. We then provide a brief conclusion in Section \ref{sec:conclusion}. 
The key result is that the inferred relation depends on how the synthetic reconstruction is selected: configurations selected using impact criteria do not recover the LMF relation, while configurations selected using the LMF discrepancy are compatible with it by construction.  

\section{Data Description} \label{sec:data}

The analyses are based on high-frequency Level 1 trade data, which provides the basic information for each executed trade, including the timestamp, traded price, traded volume, and trade direction ($\pm 1$ for buy and sell respectively). Cleaning and some pre-processing was done beforehand. Specifically, the mid-prices just before and just after the trades were included in the data sets. 

The data extended from 1 January 2023 to 31 December 2025 for the assets in Table \ref{tab:239 tickers} (See Section \ref{app:data-metadata}). To eliminate overnight order-book clearing imbalances and extreme volatility associated with the opening auction, we removed the first 10 minutes of data from each stock day. We also removed the last 10 minutes of data  for each stock day as this is when the closing auction commences. This means that we only retained data between 09:10 and 16:50. Trades with missing trade signs, quantities, prices, or other missing information were removed from the dataset. Using the data, we were able to calculate intraday volatility, daily volume, and mid-prices. This processed data was then used to generate the synthetic metaorders. Table \ref{tab:processed data sample} in Section \ref{app:data-sample processed data} has an example snippet of the data that was used.

To generate synthetic metaorders, we applied Algorithm \ref{alg:mapping}, which assigns trades to synthetic traders using a mapping function that preserves the chronological order of trades. This procedure requires specifying $N$, the number of synthetic traders, and the distribution of their trader participation, here either homogeneous or power-law. Using this "mapping", sequences of consecutive trades with the same sign for each synthetic trader are defined as metaorders and compiled using Algorithm \ref{alg:synthetic_metaorders}. Table \ref{tab:processed metaorder sample} in Section \ref{app:data-sample metaorder} has an example snippet of the output from Algorithm \ref{alg:synthetic_metaorders}.

These metaorders are subsequently used to study the square-root law of metaorder impact, the time independence of metaorder impact, the execution profile and the post execution profile. 

\section{Generating synthetic metaorders}\label{sec:synthetic}

Numerous attempts have been made to generate synthetic metaorders from public data. The validity and quality of these synthetic metaorders is tested by comparing their reproduced stylised facts to empirically proven stylised facts of metaorders. These stylised facts are \citep{Maitrieretal2025}: 
\begin{enumerate}
    \item The {\it square root law} (SQL) of metaorder impact
    \item The {\it time independence of metaorder impact}
    \item The {\it concave profile} of metaorder execution
    \item The {\it post execution decay} of metaorder impact 
\end{enumerate}
The validation of these stylised facts are individually addressed in sections \ref{subsubsec:SQL (metaorder)} and \ref{subsubsec:SQL (LMF)}, \ref{subsubsec:time independence (metaorder)} and \ref{subsubsec:time independence (LMF)}, \ref{subsubsec:execution profile (metaorder)} and \ref{subsubsec:execution profile (metaorder)},  and \ref{subsubsec:impact decay (metaorder)} and \ref{subsubsec:impact decay (LMF)} respectively. First, we will discuss the metaorder generating algorithm of \citet{Maitrieretal2025} in Section \ref{subsec:metaorder generating algorithms}.

\subsection{The metaorder generating algorithms}\label{subsec:metaorder generating algorithms}
Previous attempts have failed to reproduce all the styl\-ised facts of metaorders. As far as we know, the only method that can generate metaorders which reproduce all stylised facts is the one proposed in \citep{Maitrieretal2025}. Their algorithm uses a mapping function to assign trades to a predetermined number of traders while maintaining chronological order of the the trades. 

\begin{algorithm}[ht]
\caption{Mapping Function}
\label{alg:mapping}
\begin{algorithmic}[1]
\Require Number of traders $N$ and trader participation distribution $F$
\Ensure Trades are assigned according to participation weights $p_k$ for the $k$-th trader:
\State Draw $f_k \sim F$ for $k=1,\dots,N$
\State Normalise frequencies: $$p_k = \frac{f_k}{\sum_l f_l}$$
\State Compute cumulative probabilities $$c_k = \sum_{l=1}^{k} p_l$$ with $c_0 = 0$
\For{each order in the market}
    \State Draw a random variable $U \sim \mathcal{U}(0,1)$
    \State Find the trader $k$ such that $c_{k-1} \le U < c_k$
    \State Assign the trade to $k$-th agent.
\EndFor
\end{algorithmic}
\end{algorithm}

\begin{algorithm}[ht]
\caption{Generation of Synthetic Metaorders}
\label{alg:synthetic_metaorders}
\begin{algorithmic}[1]
\Require 
Cleaned public trade data for a given asset and trading day: trade volumes $q_{\ell}$ and $p_{\ell}$, mid-prices at order book events $\ell \in [1, N_{D}]$ for day $D$.
\Ensure 
Set of synthetic metaorders: $\{m^{(j)}_i \}_{i=1}^{n_j}$.
\State Compute daily traded volume: $\ell \in [1, N_{D}]$:
$$V_D = \sum_{\ell=1}^{N_{D}} q_{\ell}$$
\State Compute the intraday volatility:  
$$\sigma_D = \frac{m_{\max} - m_{\min}}{m_{\mathrm{open}}},$$
where $m_{\max}$ and $m_{\min}$ are the maximum and minimum mid-prices of the day, and $m_{\mathrm{open}}$ is the first non-null mid-price. Thus, $\sigma_D$ is the daily normalised high-low mid-price range used by the implementation. 
\State Apply the Mapping Function (Algorithm~\ref{alg:mapping}) to assign each trade to a synthetic trader while maintaining chronological order. 
\State Sort trades by trader and timestamp
\State Define the $j$-th metaorder as a sequence of trades with the same trade sign (from the same trader): $\{m^{(j)}_i \}_{i=1}^{n_j}$
\State Compute features of metaorder: \begin{enumerate}[label=\scriptsize{6.\arabic*}]
\item log mid-prices just before metaorder execution starts:  $\ln(p^{(j)}_0)$,
\item number of child orders: $n_j$
\item log mid-prices just after metaorder execution ends: $\ln(p^{(j)}_{n_j+1})$
\item total traded volume of the metaorder: $Q^{(j)}$
\item execution duration: $T_j$
\item any other relevant quantities.
\end{enumerate}
\State Aggregate metaorder statistics
\State Only return those with more than 1 child order
\end{algorithmic}
\end{algorithm}

We have used both homogeneous: $f \equiv f_0$, and power-law: $P(f) \propto f^{-\delta}$, distributions for our trader participation distributions. The homogeneous distribution was defined such that $f_1 = f_2 = \dots = f_N$. Samples from the power-law were obtained using inverse transform sampling. Our implementation of this algorithm allows us to specify a minimum and maximum number of trades that a trader can have, and the power-law exponent $\delta$. We have set default values of minimum = 1, maximum = total number of trades and $\delta = 2$. Additionally, we are able to specify if we want to specify a number of traders per day, per month or per year. 

Algorithm \ref{alg:synthetic_metaorders} generates the synthetic metaorders from public TAQ data. The assignment of trades to specific traders is controlled by Algorithm \ref{alg:mapping}. It is critically important that trades remain in chronological order and that the mapping is unique. Once a trade is assigned to a trader, it cannot be assigned to a different trader as well. 
Each trade receives exactly one synthetic-trader label, although the same trader label may be drawn repeatedly according to the participation probabilities \citep{Maitrieretal2025}. 

In real markets, trader activity distributions are well approximated by power-laws \citep{Maitrieretal2025}. However, the precise distribution and the optimal number of traders $N$ to generate the synthetic metaorders are not known a priori. Due to this uncertainty, we have chosen to use a power-law trader participation distribution (since it's more realistic than a homogeneous distribution) and conduct a grid search over $N$ and $\delta$. See Section \ref{sec:calibration} for details on how the grid search was conducted and Section \ref{sec:calibration results} for the results of the calibration.

\section{Calibration}\label{sec:calibration}

While Section \ref{subsec:metaorder generating algorithms} lays out the procedure for generating synthetic metaorders, we still do not know what the correct configurations \footnote{Configuration refers to the values for $N$ and $\delta$ used in Algorithm \ref{alg:mapping}} are for each stock. Intuitively, we know that not only are there a different number of active traders per day for each stock, but that the number of active traders is different each day for the same stock. Therefore, it would be wrong to use the same configuration for all the stocks for all the days. Without any prior knowledge of the number of active traders or the value of $\delta$, we use a grid search in order to find the best configurations for each stock. 

The parameter space over which we searched was $N \in$ \{5, 10, 20, 30, 40, 50, 100, 500, 1000, 1500\} and $\delta \in$ \{1.5, 2, 3, 4, 5\} traders per year, meaning, there were 50 unique configurations for each stock. We chose to use stock years instead of stock days because the stylised facts are more reliably recovered from larger amounts of data and doing configurations for each stock day would be far more computationally expensive. 

Before we can do the grid search, we first need a way to measure the quality of a configuration. We used values from three stylised facts of metaorders and a theoretical relationship from the LMF theory as targets. These are:

\begin{itemize}
    \item [1] Square root law (SQL) of metaorder impact
    \item [2] Concave execution profile of metaorders
    \item [3] Post execution impact decay of metaorders
    \item [4] $\gamma \approx \alpha - 1$ (LMF theory)
\end{itemize}

\subsection{Square Root Law (SQL)}\label{subsec:sql}

Given the $j$-th metaorder made up of $n_j$ trades, we define the total executed volume of the metaorder as:
\begin{equation}
Q^{(j)}=\sum_{i=1}^{n_j} q^{(j)}_i. \label{eq:meta-order-schedule}
\end{equation}
where $q^{(j)}_i$ is the volume of the $i$-th trade of metaorder $j$. The measured price impact of the metaorder is computed from change in the limit-order book induced by the transactions resulting from the metaorder and measured by the change in the mid-price from the start to the end of the metaorder: 

\begin{equation}\label{eq:measured metaorder impact}
    I(Q^{(j)}) = \epsilon^{(j)} \times \left(\ln(p^{(j)}_{n_j+1}) - \ln(p^{(j)}_{0})\right).
\end{equation}

Here $p_{n_j+1}$ is the first mid-price after the metaorder ends, $p_{0}$ is the last mid-price before the metaorder starts and $\epsilon^{(j)}$ is the trade sign of metaorder j. 

The SQL states that the price change induced by a large metaorder of volume $Q$ is proportional to $\sqrt{Q}$ \citep{Maitrieretal2025}. The theoretical price impact of metaorders is given as:

\begin{equation}\label{eq:meatorder impact}
    I(Q) = Y \sigma_D \sqrt{\frac{Q}{V_D}},
\end{equation}

where $Q$ is the total volume traded in the metaorder, $V_D$ is the daily traded volume, $\sigma_D$ is the intraday volatility \footnote{We used a 20 day moving average for both $V_D$ and $\sigma_D$} and $Y$ is a prefactor \citep{anomalous_price_impact}. Here for notational convenience we have dropped the metaorder and trader label; this is implicit in its definition. 

From Equation \ref{eq:measured metaorder impact} it is clear that the theoretical exponent of $Q/V_D$ is 0.5. This will be used as a target value in the grid search.

\subsection{The concave profile of metaorder impact}\label{subsec:concave profile}

The SQL gives a neat law for determining whether the executed volume $Q$ is related to the metaorder imapct $I(Q)$ in the correct way, but it does not allow us to determine whether the child orders were executed in a realistic manner. The execution profile of the metaorders are important as an analysis of 400 000 metaorders showed that there are two stages of metaorder impact, the concave profile during execution and the concave decay after execution \citep{2014marketimpactslifecycle}. In order to be realistic, the synthetic metaorders need to also have realistic execution profiles and decay. In this section we discuss the execution profiles. 

We take the the theoretical price impact profile during metaorder execution to be:
\begin{equation}\label{eq:execution profile}
    I(\phi \times Q) = \sqrt{\phi}I(Q),
\end{equation}
where $\phi \in [0, 1]$ is the fraction of the total metaorder volume that has been executed. $\phi = 0$ at the start of the metaorder and $\phi = 1$ by the end of the metaorder. More concretely, $\phi = \sum_{i = 1}^{n} q_i / Q$ and $q_i$ is the volume traded in a single child order of a metaorder of volume $Q$ \citep{Maitrieretal2025}.

Since all the metaorders have different durations, we use $\phi$ as a volume based measure of time. This has the desired effect of standardising the duration to a unit long window ($[0, 1]$). By substituting equation \ref{eq:meatorder impact} into equation \ref{eq:execution profile} we get the following:
\begin{align}
    I(\phi \times Q) &= \sqrt{\phi} \times Y \sigma_D \sqrt{\frac{Q}{V_D}}. \label{eq:execution profile (scaled)}
\end{align}
And we measured the impact profile of a metaorder as:

\begin{equation}\label{eq:measured execution profile}
    I(\phi \times Q) = \epsilon \times (\ln(p_{i+1}) - \ln(p_{0}))
\end{equation}

where $p_{0}$ is the mid-price just before the metaorder begins and $p_{i+1}$ is the mid-price just after execution of the $i^{th}$ child order. This is measured for each child order in a metaorder. 

From equations \ref{eq:execution profile (scaled)} and \ref{eq:measured execution profile}, it is clear that the impact of a portion $\phi$ of a metaorder is proportional to $\sqrt{\phi}$. Therefore, we set the target value for the exponent of $\phi$ to be 0.5. 

\subsection{Post metaorder execution impact decay}\label{subsec:post execution impact decay}

In Section \ref{subsec:concave profile} we discussed the execution profiles, next, we discuss the concave decay of price impact after execution of the metaorder. Using the same theoretical impact decay function as \citep{Maitrieretal2025}: 
\begin{equation}\label{eq:impact decay}
    I(Q,~z) = I(Q) \left(z^{1-\beta} - \left(z-1\right)^{1-\beta}\right)
\end{equation}
and after substituting in equation \ref{eq:meatorder impact} and rearranging (similarly to how we derived equation \ref{eq:execution profile (scaled)}), we get the following result:
\begin{equation}\label{eq:impact decay (scaled)}
    \frac{I(Q,~z)}{\sigma_D\sqrt{Q/V_D}} \propto \left(z^{1-\beta} - \left(z-1\right)^{1-\beta}\right)
\end{equation}
where $z = \frac{t}{T} \geq 1$ is rescaled time. $t$ is the time since the start of the metaorder $\it i.e$, $t=0$ corresponds to the start of the metaorder, $t=T$ corresponds to the end of the metaorder and $t>T$ is after the metaorder. $T$ is once again the duration of the metaorder. Both $t$ and $T$ were measured in minutes. 

The predicted value for $\beta$ is  $\frac{1-\gamma}{2}$  \citep{Maitrieretal2025}. Where $\gamma$ is the exponent for the power-law function that describes the decay of the autocorrelation of the trade signs. $\frac{1- \gamma}{2}$ is the target value for $\beta$. 

\subsection{Estimating metaorder stylised facts}

All three of the values related to the metaorder stylised facts can be estimated directly from the synthetic meta\-orders using non-linear least squares (NLLS). For each of equations \ref{eq:meatorder impact}, \ref{eq:execution profile (scaled)} and \ref{eq:impact decay (scaled)}, the fitting procedure is as follows:
\begin{enumerate}
    \item Bin the x values
    \item Fit a regression line to the data
\end{enumerate}
The results from the regressions are the fitted values for the exponents of $Q/V_D$ and $\phi$ and for $\beta$. Using NLLS also allows us to get an estimate for the variance for each fitted value respectively. 

\subsection{The LMF theory}\label{sec:lmf theory}

The Lillo-Mike-Farmer (LMF) theory provides a microscopic explanation for the long-range correlation (LRC) observed in market-order flow. Empirically, trade signs of market orders exhibit strong persistence. The LMF model attributes this macroscopic long memory to the microscopic behaviour of order-splitting traders (STs) who divide large orders into sequences of smaller child orders of the same sign (metaorders) \citep{SatoKanazawa2023LMFTest}.

\subsection{Empirical Estimation of $\alpha$ and $\gamma$}\label{subsec:emstimation of alpha and gamma}

The key assumption of the LMF theory is that meta\-order run lengths $L$ \footnote{Run lengths $L$ refers to the number of child orders in a metaorder.} follow a power-law distribution:
\begin{equation}\label{eq:run length distribution}
P(L) \sim L^{-\alpha-1}, \quad \alpha > 1
\end{equation}
and that the presence of STs is sufficient to generate the observed LRC. Specifically, the LMF theory suggests that a direct consequence of the power-law distribution of $L$ is that the LRC of order-flow decays as a power-law \citep{LilloFarmer2004}:
\begin{equation}\label{eq:power-law decay}
    C(\tau) \propto \tau^{-\gamma}
\end{equation}
Where $C(\tau)$ is the autocorrelation at lag $\tau$ and $\gamma$ is the power-law exponent. 

Under the assumption of random order submissions by STs, the model predicts a simple quantitative relation between the microscopic exponent $\alpha$ and the macroscopic LRC exponent $\gamma$:

\begin{equation}\label{eq:alpha gamma}
\gamma \approx \alpha - 1.
\end{equation}

This remarkable result connects a directly measurable macroscopic property of the market (the decay of order sign autocorrelations) to a microscopic property of trader behaviour (the distribution of metaorder run lengths). Eq\-uation \ref{eq:alpha gamma} gives us a target value for $\alpha$, we want it to be $\gamma + 1$.

In order to use this target we have to estimate both $\gamma$ and $\alpha$. The details of the steps followed to estimate them are outlined below. 

\subsubsection{Strategy clustering}

To test this prediction, we first classify traders into STs and random traders (RTs) using a strategy clustering procedure. We used a one-sided runs test to test for clustering. A run is defined as a sequence of consecutive same-sign orders. The null hypothesis for the test is that the trader is a RT and the alternate hypothesis is that the trader is a ST. We classify a trader as an ST if they have a statistically significant lower number of runs than a trader who randomly buys or sells would have. This is a one-sided test since traders who have more runs than a trader randomly buying or selling are also classified as RTs. This was done because having more runs is indicative of behaviour more akin to alternating between buying a selling than it is to making sequences of buys or sells. $L$ is then extracted from the orders of STs.

\subsubsection{Autocorrelations}
In order to  show this, we start by defining the sample autocorrelation (in the time domain) as:

\begin{equation}\label{eq:sample autocorrelation}
\hat{C}(\tau) \coloneq \sum_{\ell=1}^{N_{\epsilon}-\tau} \frac{\epsilon_{\ell}\epsilon_{\ell+\tau}}{N_{\epsilon} - \tau},
\end{equation}

where $\epsilon_{\ell} = \pm 1$ are the measured market-order signs and $N_{\epsilon}$ is the number of market orders \citep{SatoKanazawa2023LMFTest}. 
We compute this uncentred sign-product estimator using an FFT zero-padded to the next power of two at least as large as $2N_{\epsilon}-1$. The lagged sum at lag $\tau$ is divided by $N_{\epsilon}-\tau$, which prevents circular wraparound over the retained lags and gives the normalisation in Equation \ref{eq:sample autocorrelation}.

Calculating autocorrelations in the time domain is computationally expensive, particularly for $\tau$. Therefore, we use the convolution theorem and the Fast Fourier Transform (FFT) to calculate the autocorrelations instead. We do this

\begin{equation}\label{eq:sample autocorrelation (FFT)}
\hat{C}(\tau) = \frac{\mathcal{F}^{-1}\left[\mathcal{F}(\epsilon) \mathcal{F}^{*}(\epsilon)\right]_\tau}{N_{\epsilon} - \tau}
\end{equation}

Where $\mathcal{F}$ is the Fourier transform and $\mathcal{F}^{*}$ is the complex conjugate of $\mathcal{F}$ \citep{numericalrecipes}. 

\subsubsection{Choosing an appropriate range}

For large $\tau$, the autocorrelations decay and become noise. It is therefore important that we choose an appropriate range over which to fit the power-law so that our estimate of $\gamma$ is not affected by noise in the tails. 

We start by first truncating the sequence of autocorrelations at the first instance where it is noise. We then identified the best range for fitting the power-law by first performing a log transform, thereby making the decay linear

\begin{equation}\label{eq:power-law decay (log scaled)}
    \mathrm{ln}(C(\tau)) \propto -\gamma  \mathrm{ln}(\tau)
\end{equation}

and allowing us to use linear regression and the r-squared statistic to compare candidate tail ranges. After truncation, the implementation fixes the right endpoint and progressively increases the left endpoint, subject to retaining at least 10,000 points. It calculates the log-log r-squared statistic for each candidate and uses the range with the highest value.

\subsubsection{Estimating $\gamma$ and $\alpha$}

After the fitment range was decided, we then estimate $\gamma$.  We used NLLS and power-spectral density (PSD). The NLLS procedure is the same as for the metaorder stylised facts. We used the log transformed values from Equation \ref{eq:power-law decay (log scaled)} to make the fitting procedure more stable. 

For the spectral estimate, we take the absolute FFT amplitude of the estimated autocorrelation sequence, retain positive frequencies, and fit the lowest 15\% of these frequencies in log-log coordinates. Writing the fitted relation as $\log S(f)=a+s\log f$, we estimate $\widehat{\gamma}=s+1$. This is an ACF-based spectral estimate rather than a periodogram fitted directly to the trade-sign sequence. 

We fit the discrete run-length distribution using \texttt{power\-law.Fit}, with the lower cutoff $L_{\min}$ selected by the package \citep{Clausetetal2009}. If the package returns the probability-mass exponent $a$ in $P(L)\propto L^{-a}$, we report $\alpha=a-1$, so that the manuscript convention is $P(L)\propto L^{-\alpha-1}$. No separate bootstrap goodness-of-fit test is used in the committed implementation.

\subsection{Summary of calibration targets}

Here we now summarise the calibration targets. For each combination of $N ~ \mathrm{and} ~ \delta$ we generated the metaorders and then estimated the exponent of $Q/V_D$ (Equation \ref{eq:meatorder impact}), the exponent of $\phi$ (Equation \ref{eq:execution profile (scaled)}), $\beta$ (Equation \ref{eq:impact decay (scaled)}), $\alpha$ (Equation \ref{eq:run length distribution}) and $\gamma$ (Equation \ref{eq:power-law decay}) for each stock year, when using the NLLS and PSD methods. The parameters to be estimated and their target values are summarised in Table \ref{tab:estimated parameters}

\begin{table}[H]
    \centering
    \begin{tabular}{llc}
    \toprule
         & parameter & theoretical value \\
         \midrule
         \multirow{3}{*}{\makecell{metaorder stylised\\ facts}} & $Q/V_D$ exponent & 0.5 \\
         & $\phi$ exponent             & 0.5 \\
         & $\beta$                     & $\frac{1-\gamma}{2}$ \\
         \midrule 
         \multirow{2}{*}{LMF theory} 
         & $\alpha$                    & $\gamma + 1$ \\
         & $\gamma$                    & - \\
    \bottomrule
    \end{tabular}
    \caption{The parameters estimated for each configuration of the grid search and the target value of each. There is no theoretical value for $\gamma$ since it is measured directly from the autocorrelation of the trade signs.}
    \label{tab:estimated parameters}
\end{table}

\section{Calibration results}\label{sec:calibration results}

The objective of the grid search is to find parameter configurations that align the stylised facts and LMF results of the synthetic metaorders to the empirical results of real metaorders. While the parameters estimated and their respective targets are summarised in Table \ref{tab:estimated parameters}, we still need a way to quantify the quality of the configurations. We start with defining how the error between estimate and target is calculated.   

The following variance-aware loss function was used to quantify the error for each parameter estimate relating to the metaorder stylised facts: 

\begin{equation}\label{eq:grid search metaorder loss function}
    \mathrm{e} = \frac{|\hat{\theta} - \theta|}{\theta} + \lambda \times \frac{\mathrm{VAR}(\hat{\theta})}{\theta ^{2}}
\end{equation}

Where $\hat{\theta}$ is the point estimate, $\theta$ is the theoretical value of the estimate and $\lambda$ is the coefficient of the penalty term. By imposing the penalty on the variance, we are selecting for both accuracy and for low variance. 

The individual errors for the parameters were then aggregated into a single value:

\begin{equation}\label{eq:grid search metaorder aggregated error}
    \mathrm{e}_{M} = \frac{\sum_{i=1}^{3}\eta_i  \mathrm{e}_i}{\sum_{i=1}^3 \eta_i}
\end{equation}

Where $\mathrm{e}_M$ is the aggregated error, $\mathrm{e}_i$ are the individual errors and $\eta_i$ are the coefficients for each of the error terms. We used $\eta_{i} = 1$ for $i = 1,2,3$ as the default, however, the specific values can be altered to reflect placing greater emphasis on minimising the error for a specific stylised fact. \footnote{For example, setting $\eta_1 = 2$ and $\eta_{2,3} = 1$ means that the error for the exponent of $Q/V_D$ contributes more to $\mathrm{e}_M$} 

For the estimate of $\alpha$, we did not measure the variance of the estimate and as such we minimised a simpler loss function:

\begin{equation}\label{eq:grid search LMF loss function}
    \mathrm{e}_{LMF} = \frac{|\hat{\alpha} - \hat{\gamma} - 1|}{\hat{\gamma} + 1}
\end{equation}

Using equations \ref{eq:grid search metaorder loss function}, \ref{eq:grid search metaorder aggregated error} and \ref{eq:grid search LMF loss function}, we are able to choose configurations for each stock year that minimise either $\mathrm{e}_{M}$ or $\mathrm{e}_{LMF}$. 
$\mathrm{e}_M$ measures discrepancy from the selected impact targets, while $\mathrm{e}_{LMF}$ measures in-sample discrepancy from the LMF exponent relation. When $\mathrm{e}_{LMF}$ is used to select a configuration, the resulting agreement is not an independent validation of the relation. Having both quantities allows us to investigate how optimising for $\mathrm{e}_M$ affects $\mathrm{e}_{LMF}$ and vice-versa. 

\subsection{Minimising \texorpdfstring{$\mathrm{e}_M$}{Metaorder stylised facts error}}\label{subsec:minimising e_M}

\begin{table*}[t]
\centering
\begin{tabular}{lcccccccccc}
\toprule
ticker & year & N & $\delta$ & $\alpha$ & $\gamma$ & \makecell{$Q/V_D$ \\ exponent} & \makecell{$\phi$ \\ exponent} & $\beta$ & $\mathrm{e}_{M}$ (PSD) & $\mathrm{e}_{LMF}$ (PSD) \\
\midrule
GFI & 2023 & 50 & 5.00 & 1.62 & 0.57 & 0.04 & 0.91 & 0.17 & 0.66 & 0.03 \\
GFI & 2024 & 100 & 4.00 & 1.67 & 0.56 & 0.04 & 0.84 & 0.17 & 0.60 & 0.07 \\
GFI & 2025 & 100 & 3.00 & 1.68 & 0.64 & 0.04 & 0.82 & 0.16 & 0.55 & 0.02 \\
\bottomrule
\end{tabular}
\caption{Best configurations for GFI when selecting for optimal metaorder stylised fact results (minimising $\mathrm{e}_{M}$), using power-laws for the trader participation distributions and using the PSD method to estimate $\gamma$. We used $\lambda$ = 1 (equation \ref{eq:grid search metaorder loss function}) and $\eta_i$ = 1 (equation \ref{eq:grid search metaorder aggregated error}).}
\label{tab:optimal configs GFI (metaorder) (PSD) (power)}
\end{table*}

\begin{table*}[t]
\centering
\begin{tabular}{lrrrrrrrrrr}
\toprule
ticker & year & N & $\delta$ & $\alpha$ & $\gamma$ & \makecell{$Q/V_D$ \\ exponent} & \makecell{$\phi$ \\ exponent} & $\beta$ & $\mathrm{e}_{M}$ (PSD) & $\mathrm{e}_{LMF}$ (PSD) \\
\midrule
GRT & 2023 & 100 & 4.00 & 1.62 & 0.48 & 0.15 & 0.78 & 0.24 & 0.45 & 0.09 \\
GRT & 2024 & 1500 & 3.00 & 1.60 & 0.43 & 0.02 & 0.56 & 0.19 & 0.49 & 0.12 \\
GRT & 2025 & 100 & 5.00 & 1.62 & 0.49 & 0.13 & 0.69 & 0.15 & 0.51 & 0.09 \\
\bottomrule
\bottomrule
\end{tabular}
\caption{Best configurations for GRT when selecting for optimal metaorder stylised fact results (minimising $\mathrm{e}_{M}$), using power-laws for the trader participation distributions and using the PSD method to estimate $\gamma$. We used $\lambda$ = 1 (equation \ref{eq:grid search metaorder loss function}) and $\eta_i$ = 1 (equation \ref{eq:grid search metaorder aggregated error}).}
\label{tab:optimal configs GRT (metaorder) (PSD) (power)}
\end{table*}

Tables \ref{tab:optimal configs GFI (metaorder) (PSD) (power)} and \ref{tab:optimal configs GRT (metaorder) (PSD) (power)} reports the optimal configurations for GFI and GRT when minimising $\mathrm{e}_{M}$, using a power-law trader participation distribution and using the PSD method to estimate $\gamma$. We have used a penalty coefficient of $\lambda = 1$ for each stylised fact and used $\eta_i = 1$ for each of the error terms. 


\subsubsection{SQL ($\mathrm{e}_M$)}\label{subsubsec:SQL (metaorder)}

\begin{figure*}[h!]
    \centering
    \begin{subfigure}[h]{0.45\linewidth}
        \includegraphics[width=\textwidth]{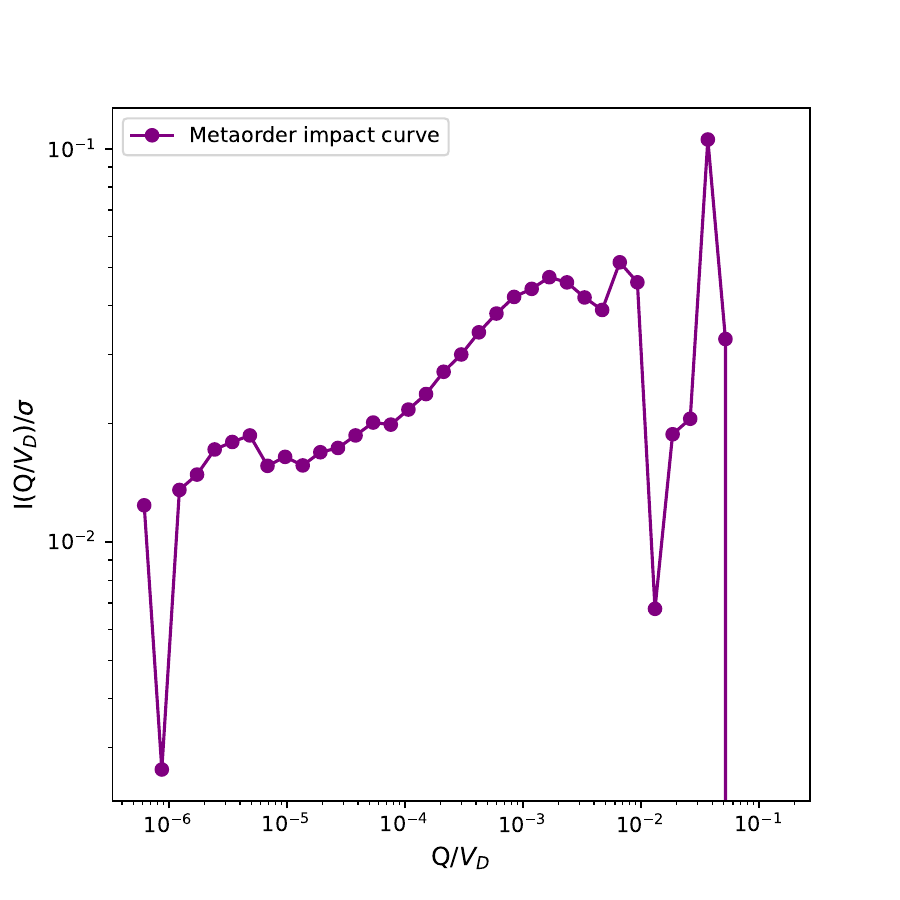}
        \caption{GFI}
        \label{subfig:calibrated GFI impact curve (metaorder)}
    \end{subfigure}
    \begin{subfigure}[h]{0.45\linewidth}
        \includegraphics[width=\textwidth]{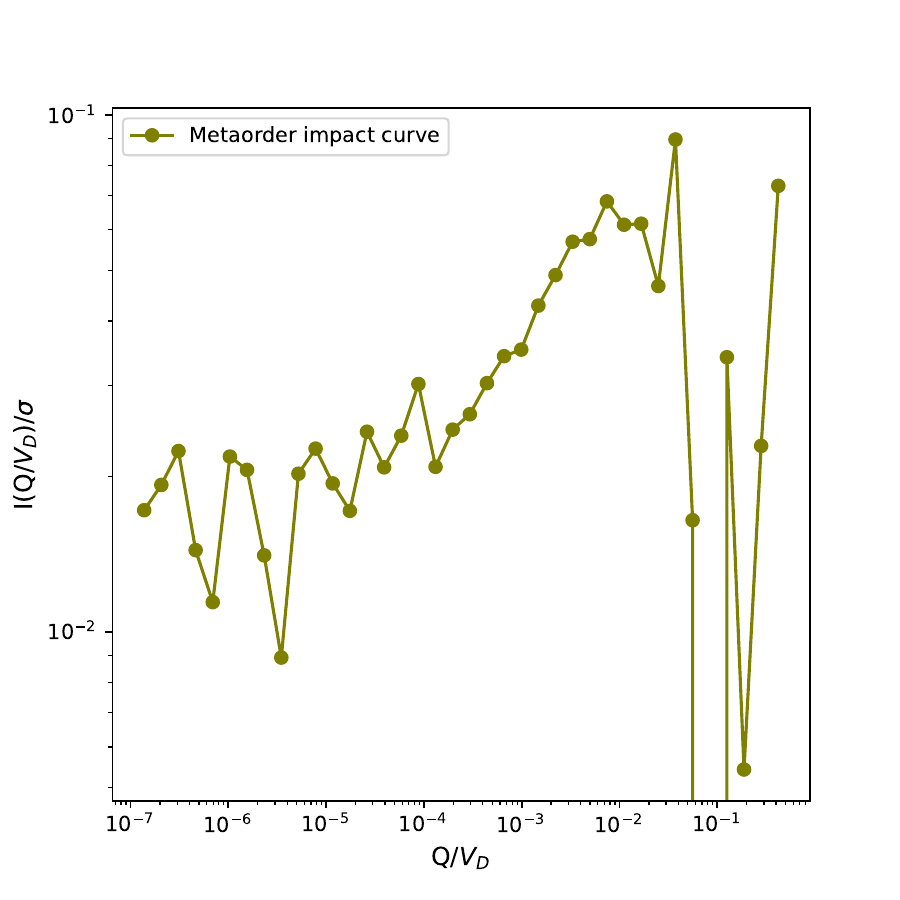}
        \caption{GRT}
        \label{subfig:calibrated GRT impact curve (metaorder)}
    \end{subfigure}
    \caption{Metaorder impact curves for GFI (a) and GRT (b) using stock year configurations that minimised $\mathrm{e}_M$ (when using the PSD method to estimate $\gamma$). Power-laws were used for the trader participation distributions. See tables \ref{tab:optimal configs GFI (metaorder) (PSD) (power)} and \ref{tab:optimal configs GRT (metaorder) (PSD) (power)} for the exact configurations used for each stock year for GFI and GRT respectively. It is clear that the impact curves for both GFI (a) and GRT (b) have noisy tails due to there being few observations in those bins. For GRT (b), the curve is also noisy. However, when metaorders for all stocks are pooled and aggregated (Figure \ref{fig:calibrated market impact curve (metaorder)}), the curve has much less noise.}
    \label{fig:calibrated impact curves (metaorder)}
\end{figure*} 

The metaorder impact curves for GFI and GRT when minimising $\mathrm{e}_M$ and using the PSD method to estimate $\gamma$ are shown in Figure \ref{fig:calibrated impact curves (metaorder)}. Both axes are logarithmically scaled. The y-axis is the metaorder impact, scaled by $\frac{1}{\sigma}$. Here, $\sigma$ is the 20 day average intraday volatility. The x-axis is the volume of the metaorder, scaled by $\frac{1}{V_D}$. $V_D$ is the 20 day average volume. Scaling the metaorder volume by the 20 day average. 

The plots were created by binning the observations into 40 equally spaced bins (in logarithmic space) according to their $Q/V_D$ value and using the bin average scaled impact as the representative value for each bin. The bin centres and their representative points are then plotted. From the Figure it is immediately clear that for both large and small $Q/V_D$ there is a significant amount of noise. This is due to there being very few observations in these bins so there is significant bias in the mean of these bins.

\begin{figure}[h!]
    \centering
    \includegraphics[width=0.5\textwidth]{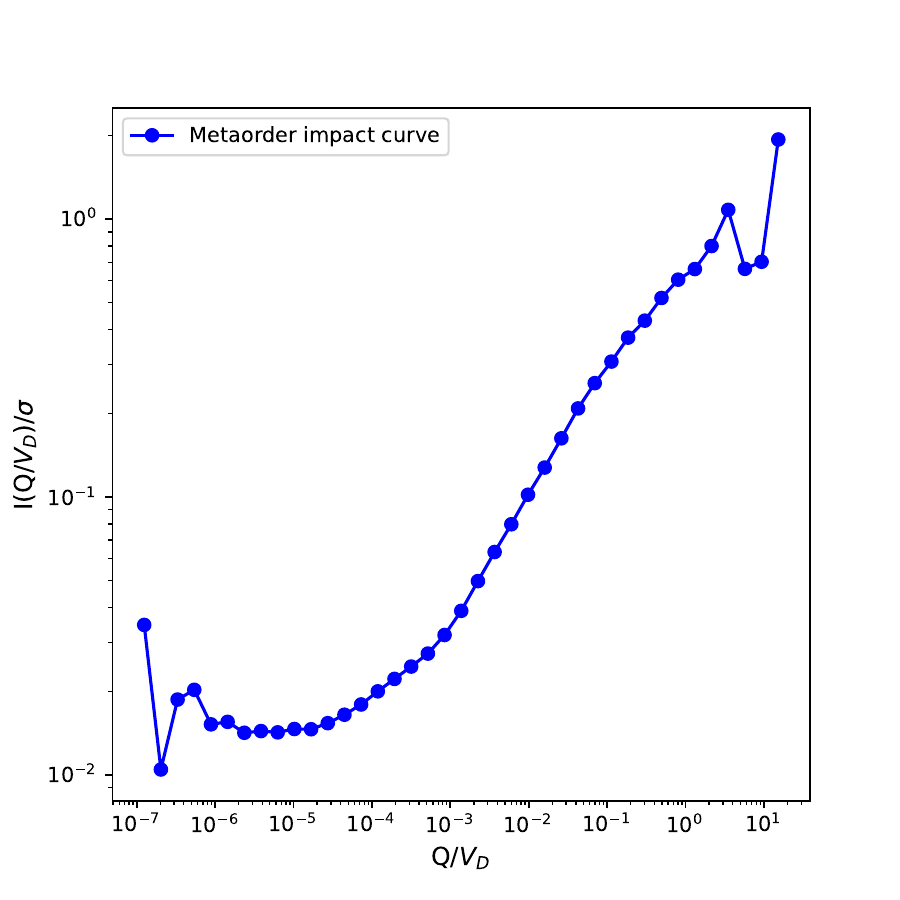}
    \caption{Metaorder impact curve for our entire universe of stocks, pooled and aggregated into a single curve, when using stock year configurations that minimised $\mathrm{e}_M$ (when using the PSD method to estimate $\gamma$). Power-laws were used for the trader participation distributions. When pooled and aggregated, the impact curve obeys the SQL. Notably, there is considerably less noise in the tails compared to Figure \ref{fig:calibrated impact curves (metaorder)} and we also see that for small relative metaorders $\left(Q/V_D \leq 10^{-2}\right)$ there is a plateau where the impact is larger than the SQL predicts. This result is similar to the results found in \citep{Harvey_2017}.}
    \label{fig:calibrated market impact curve (metaorder)}
\end{figure} 

\begin{table*}[h!]
\centering
\begin{tabular}{cccccc}
\toprule
\makecell{trade \\sign} & count & \makecell{proportion of \\data} & \makecell{average \\ $Q/V_D$  \\ $[\times 10^{-3}]$} & \makecell{proportion of metaorders \\ with $Q/V_D \leq 10^{-2}$} & \makecell{average $Q/V_D$ for metaorders \\ with $Q/V_D \leq 10^{-2}$ \\ $[\times 10^{-4}]$} \\
\midrule
1 & 24596453 & 50.04 & 1.464 & 39.53 & 3.47 \\
-1 & 24561605 & 49.96 & 1.470 & 39.41 & 3.47 \\
\bottomrule
\end{tabular}
\caption{Checking for trade sign imbalance in synthetic metaorders when using configurations that minimised $\mathrm{e}_{M}$. The data is split approximately equally between buy and sell metaorders, trade signs of 1 and -1 respectively. The average relative size of the metaorders, $Q/V_D$ is nearly identical ($1.464 \times 10^{-3}$ and $1.470 \times 10^{-3}$ respectively) between buy and sell metaorders. So it is clear that the data is not imbalanced and there is no meaningful difference in the relative sizes of the buy and sell metaorders respectively. From Figure \ref{fig:calibrated market impact curve (metaorder)}, the plateau appears for $Q/V_D \leq 10^{-2}$ but it is clear that the proportion of metaorders with $Q/V_D \leq 10^{-2}$ is approximately the same for buy and sell metaorders. Additionally, the average $Q/V_D$ for that proportion is also identical ($3.47 \times 10^{-4}$) for buy and sell metaorders. These results show that a trade sign imbalance in the dataset is not causing the plateau in Figure \ref{fig:calibrated market impact curve (metaorder)}.}
\label{tab:trade sign imbalance table (metaorder)}
\end{table*}

Figure \ref{fig:calibrated market impact curve (metaorder)} shows the metaorder impact curve for the entire universe of stocks, aggregated into a single curve. Metaorders were generated for each stock year by using the configuration that minimised $\mathrm{e}_M$ (when using the PSD method to estimate $\gamma$) for that stock year. The aggregation was done by pooling the metaorders for each stock into a single data set containing all the metaorders in the entire market. These were then plotting in the same way as Figure \ref{fig:calibrated impact curves (metaorder)}. 

From Figure \ref{fig:calibrated market impact curve (metaorder)} it is clear that on aggregate, the meta\-orders obey the SQL for $Q/V_D \geq 10^{-2}$. However, for $Q/V_D < 10^{-2}$ the metaorder impact flattens significantly and forms a plateau. One possible explanation for this plateau is a buy and sell imbalance in the synthetic meta\-orders, particularly for $Q/V_D \leq 10^{-2}$. To test this, Table \ref{tab:trade sign imbalance table (metaorder)} compares the number of meta\-orders, the porportion of data, the average $Q/V_D$, proportion of metaorders with $Q/V_D \leq 10^{-2}$ and the average $Q/V_D$ for metaorders with $Q/V_D \leq 10^{-2}$ for buy and sell synthetic metaorders. The metaorders were constructed using configurations thats minimised $\mathrm{e}_{M}$ From the table, the buy and sell metaorders are almost perfectly balanced (50.04\% and 49.96\% respectively), while the average relative metaorder sizes $Q/V_D$ were nearly identical ($1.464 \times 10^{-3}$ and $1.470 \times 10^{-3}$ respectively). Table \ref{tab:trade sign imbalance table (metaorder)} also shows that there is no imbalance in the proportion of buy and sell metaorders that have relative sizes $Q/V_D \leq 10^{-2}$ (39.53\% and 39.41\% respectively) and the average relative size $Q/V_D$ of this proportion is identical ($3.47 \times 10^{-4}$) for both buy and sell metaorders. The results in the table show that the plateau is unlikely to be caused by a trade sign imbalance in the synthetic metaorders. Instead, the observed plateau is consistent with the results found in \citep{Harvey_2017}, who report that small transactions exhibit larger-than-expected price impact after the JSE underwent a fee restructuring on 30-Sep-2013.

\subsubsection{Time independence ($\mathrm{e}_M$)}\label{subsubsec:time independence (metaorder)}

It has been shown that the impact of a metaorder is independent of its duration \citep{Maitrieretal2025}.  

\begin{figure*}[h!]
    \centering
    \begin{subfigure}[h]{0.45\linewidth}
        \includegraphics[width=\textwidth]{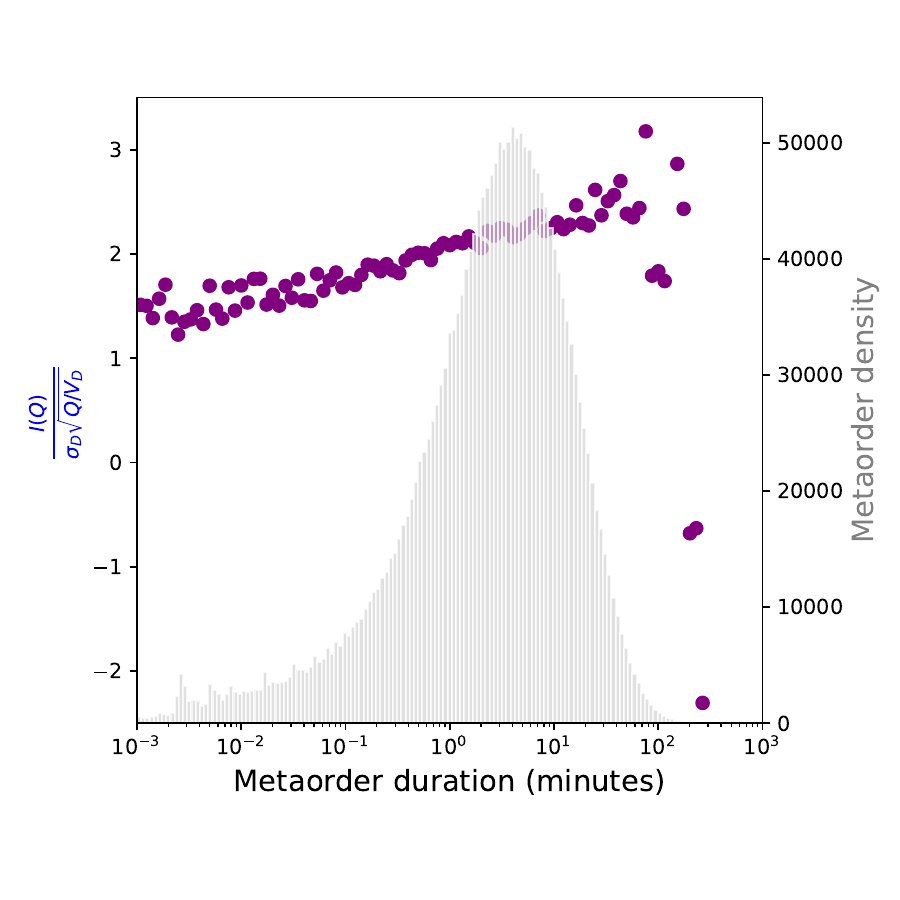}
        \caption{GFI}
        \label{subfig:calibrated GFI time independence (metaorder)}
    \end{subfigure}
    \begin{subfigure}[h]{0.45\linewidth}
        \includegraphics[width=\textwidth]{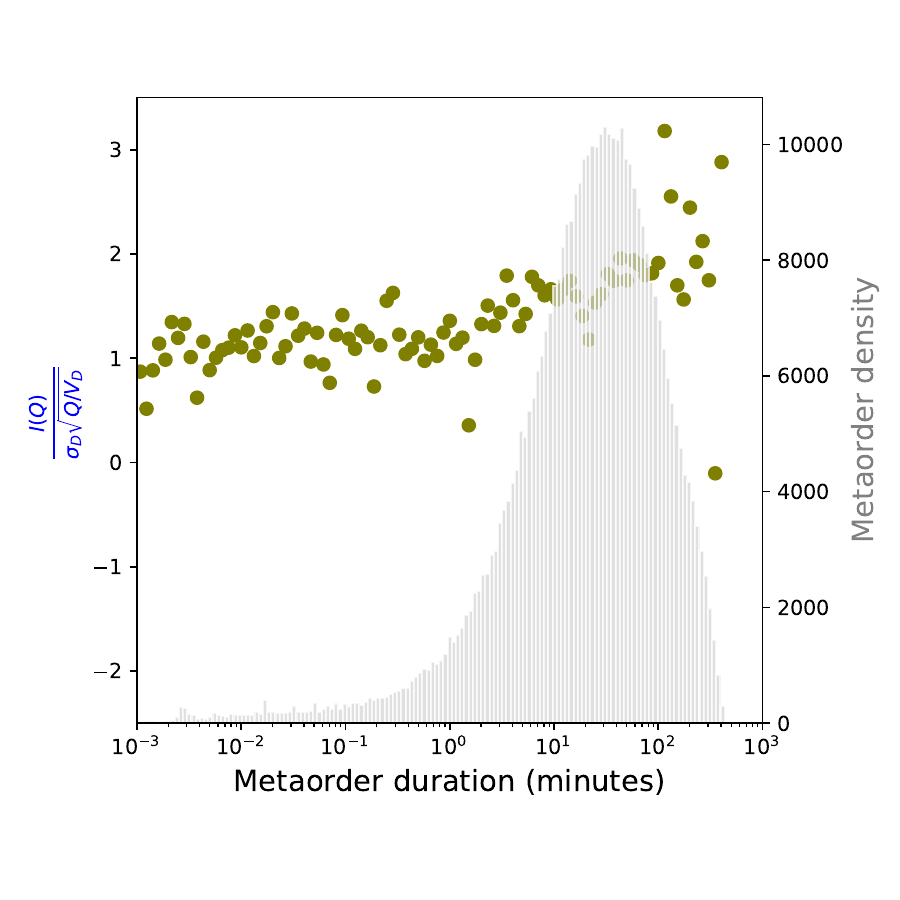}
        \caption{GRT}
        \label{subfig:calibrated GRT time independence (metaorder)}
    \end{subfigure}
    \caption{Time independence of metaorder impact plots for GFI (a) and GRT (b) when using stock year configurations that minimised $\mathrm{e}_M$ (when using the PSD method to estimate $\gamma$). Power-laws were used for the trader participation distributions. For GFI, the average metaorder duration is approximately $10^{0.75}$ (5) minutes while for GRT the average duration is approximately $10^{2}$ (100) minutes. This longer duration for GRT may be due to there being less liquidity, meaning metaorders take longer to execute fully. Although there could be a number of other factors that could contribute to this artifact.}
    \label{fig:calibrated time independence (metaorder)}
\end{figure*} 

Figure \ref{fig:calibrated time independence (metaorder)} shows the independence of metaorder impact with respect to the metaorder duration for GFI and GRT when using metaorder configurations that minimise $\mathrm{e}_M$, power-law trader participation distributions and the PSD method to estimate $\gamma$. A histogram of the metaorder durations is given in grey and the dynamic impact is overlayed. The x and y axes of the histogram are the metaorder durations and density respectively. The x and y axes of the overlayed plot are the metaorder durations and dynamic impact respectively. In both cases, the x axis is logarithmically scaled.

The overlayed plots were created by binning the observations into 100 equally spaced bins (in logarithmic space) according to their duration and using the average dynamic impact as the representative value for each bin. The bin centers and their representative points are then plotted. Given the logarithmically scaled x-axis, it is clear that the impact remains relatively flat as the metaorder duration increases. This indicates approximate independence between the impact and the duration of the metaorders. This result is consistent with the results found in \citep{Maitrieretal2025}. The average metaorder duration for GFI (Figure \ref{subfig:calibrated GFI time independence (metaorder)}) is slightly less than 10 minutes, and for GRT (Figure \ref{subfig:calibrated GRT time independence (metaorder)}) it is slightly less than 100 minutes. Both sub-figures show long left tails with short right tails, indicating far more metaorders with short durations compared to long durations. This could indicate the presence of high frequency trading. 

\subsubsection{Execution profile ($\mathrm{e}_M$)}\label{subsubsec:execution profile (metaorder)}

\begin{figure*}[h!]
    \centering
    \begin{subfigure}[h]{0.45\linewidth}
        \includegraphics[width=\textwidth]{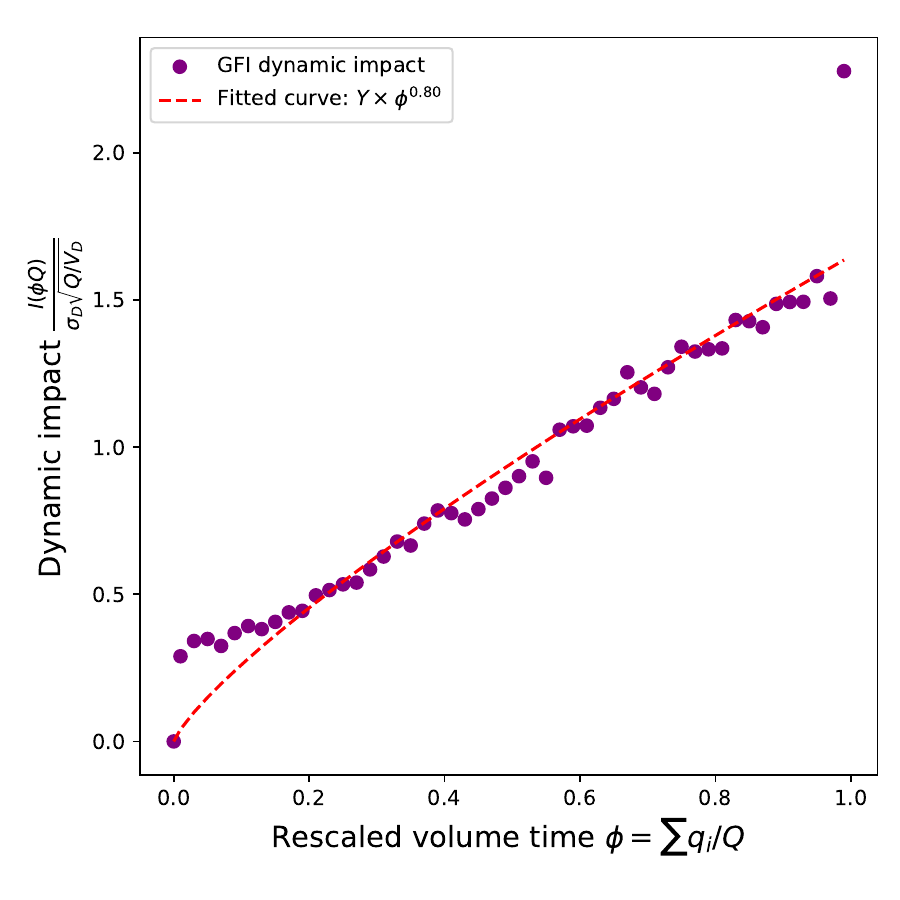}
        \caption{GFI}
        \label{subfig:calibrated GFI concave profile (metaorder)}
    \end{subfigure}
    \begin{subfigure}[h]{0.45\linewidth}
        \includegraphics[width=\textwidth]{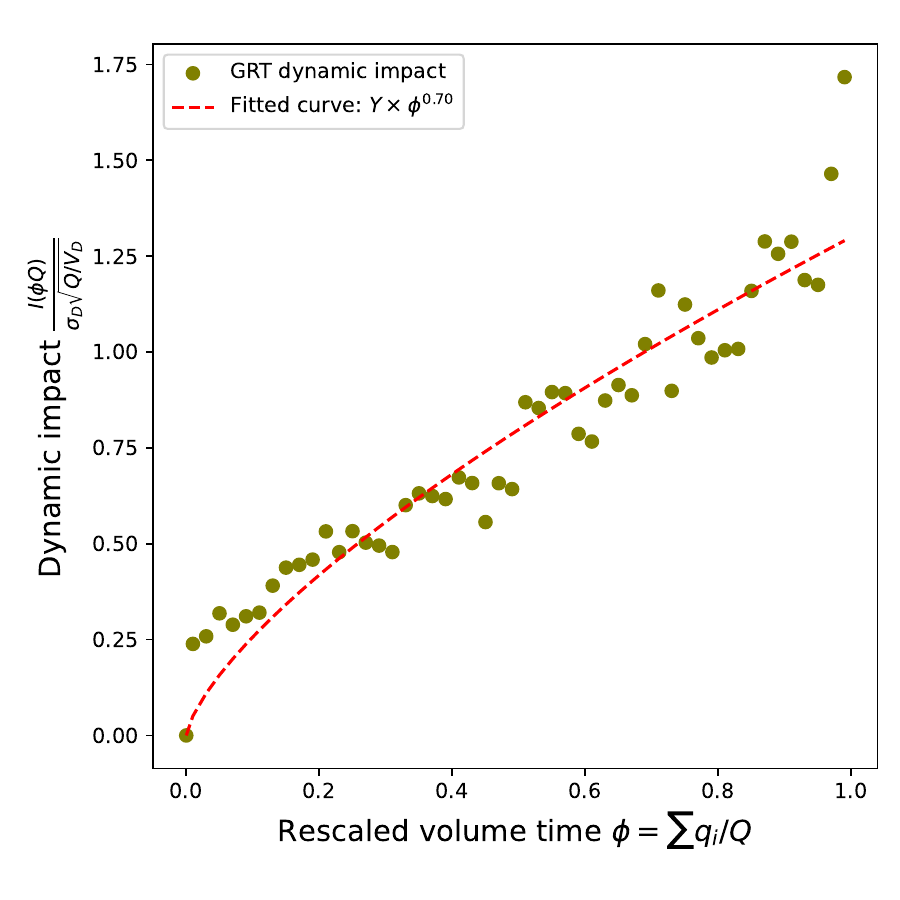}
        \caption{GRT}
        \label{subfig:calibrated GRT concave profile (metaorder)}
    \end{subfigure}
    \caption{Concave profile of metaorder impact for GFI (a) and GRT (b) when using stock year configurations that minimised $\mathrm{e}_M$ (when using the PSD method to estimate $\gamma$). Power-laws were used for the trader participation distributions. Only metaorders with 3 or more child orders were used to create these plots. The dynamic impact is plotted as a function of $\phi$, the proportion of the metaorder which has been executed. The fitted curves are also plotted in red. See tables \ref{tab:optimal configs GFI (metaorder) (PSD) (power)} and \ref{tab:optimal configs GRT (metaorder) (PSD) (power)} for the exact configurations used for each stock year for GFI and GRT respectively. See Table \ref{tab:calibrated concave profiles (metaorder)} for the fitted values for the fited curves of GFI and GRT respectively.}
    \label{fig:calibrated concave profiles (metaorder)}
\end{figure*} 

Figure \ref{fig:calibrated concave profiles (metaorder)} shows the execution profiles for GFI and GRT when minimising $\mathrm{e}_M$ and using the PSD method to estimate $\gamma$, with the fitted curves shown as a dashed red line. The y-axis is the dynamic impact and the x-axis is the proportion of the metaorder that has been executed, check equation \ref{eq:execution profile (scaled)}. Only metaorders with 3 or more child orders were included. 

The plots were created by binning the observations into 50 equally spaced bins based on $\phi$ value and then taking the average dynamic impact as the representative point for each bin. The bin centres and their representative points are then plotted. It is clear that there is more noise in the points for GRT (Figure \ref{subfig:calibrated GRT concave profile (metaorder)}) than for GFI (Figure \ref{subfig:calibrated GFI concave profile (metaorder)}). This may however be due to there being too few observations per bin as a result of there being much fewer metaorders for GRT compared to GFI. Despite this, the fits in both cases are good and there are no clear deviations from the points. Table \ref{tab:calibrated concave profiles (metaorder)} reports the fitted values with the 95\% confidence interval half widths for the fitted curves in figures \ref{subfig:calibrated GFI concave profile (metaorder)} and \ref{subfig:calibrated GRT concave profile (metaorder)}. 

\begin{table}[h!]
\centering
\begin{tabular}{lccc}
\toprule
Description & Ticker & $Y$ & \makecell{$\phi$ \\ exponent} \\
\midrule
\makecell{Gold Fields \\ Ltd.} & GFI & 1.649 $\pm$ 0.039 & 0.802 $\pm$ 0.044\\
\makecell{Growthpoint \\ Ltd.} & GRT & 1.300 $\pm$ 0.035 & 0.704 $\pm$ 0.046\\ 
\bottomrule
\end{tabular}
\caption{Fitted values for metaorder execution profiles of GFI and GRT. These are the fitted values for figures \ref{subfig:calibrated GFI concave profile (metaorder)} and \ref{subfig:calibrated GRT concave profile (metaorder)}.}
\label{tab:calibrated concave profiles (metaorder)}
\end{table}

From the Table it is clear that the exponents of $\phi$ for both GFI and GRT are greater than the theoretical value of 0.5. Specifically, the exponent of $\phi$ for GFI is $0.802 \pm 0.044$ and for GRT the exponent is $0.704 \pm 0.046$. In both cases, the confidence interval half widths are narrow, which indicates that the fitted values are precise. 

\subsubsection{Impact decay ($\mathrm{e}_M$)}\label{subsubsec:impact decay (metaorder)}

\begin{figure*}[h!]
    \centering
    \begin{subfigure}[h]{0.45\linewidth}
        \includegraphics[width=\textwidth]{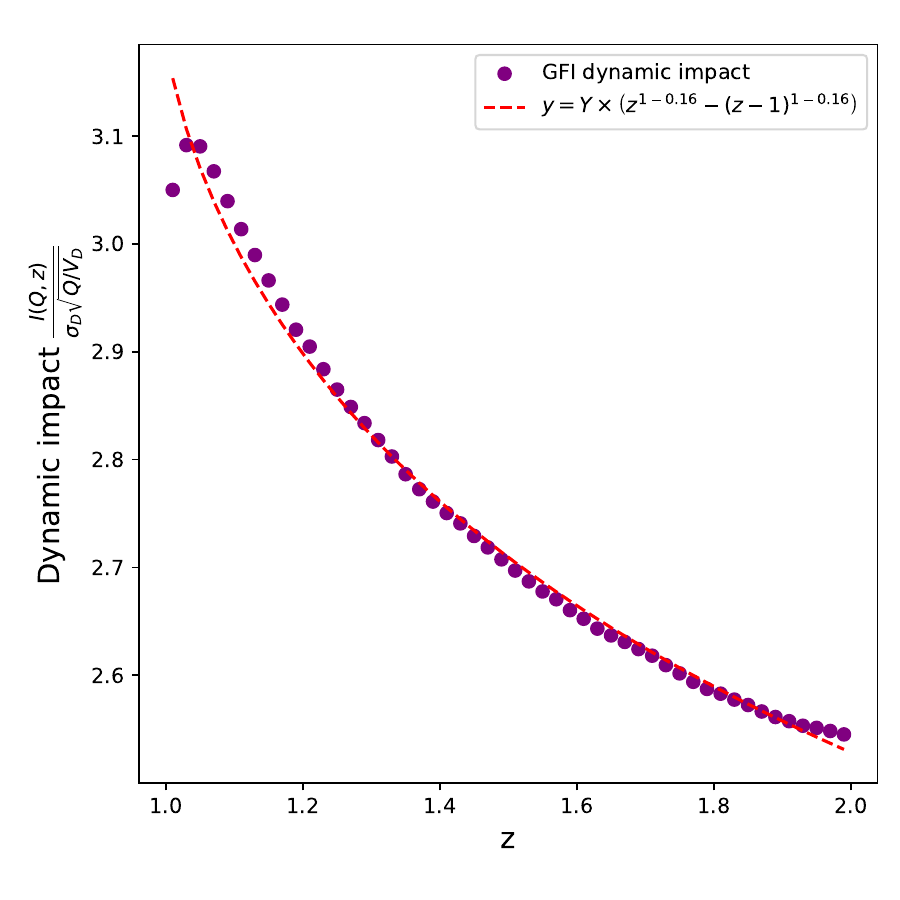}
        \caption{GFI}
        \label{subfig:calibrated GFI impact decay (metaorder)}
    \end{subfigure}
    \begin{subfigure}[h]{0.45\linewidth}
        \includegraphics[width=\textwidth]{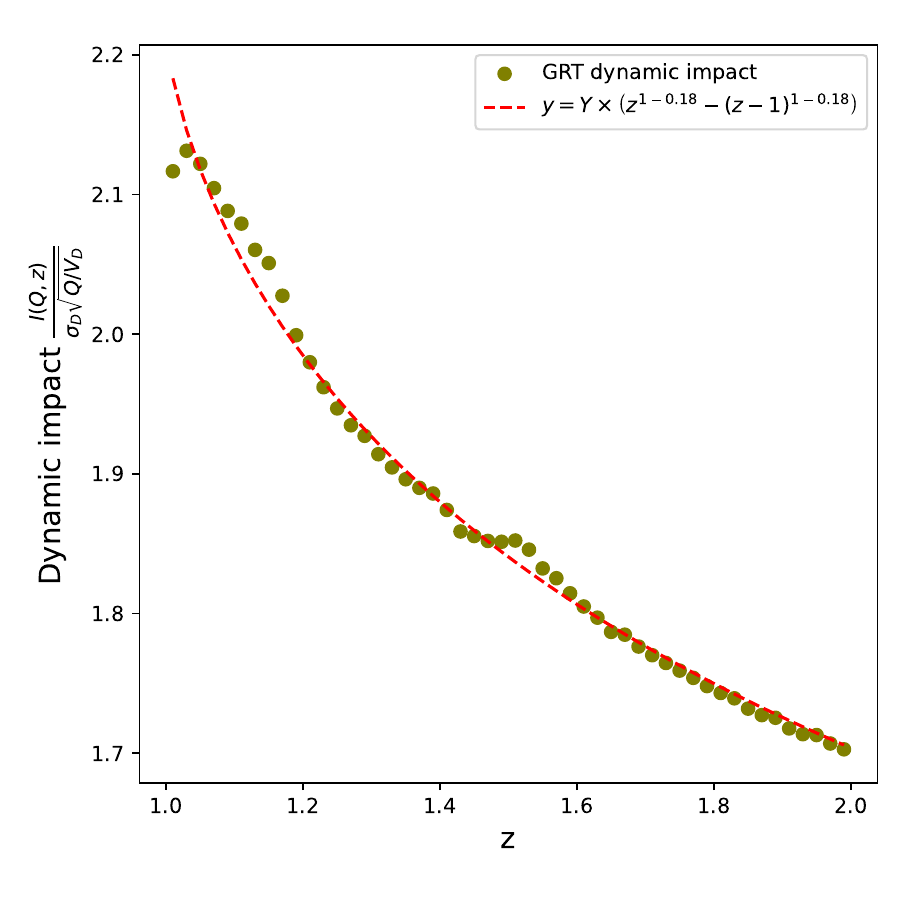}
        \caption{GRT}
        \label{subfig:calibrated GRT impact decay (metaorder)}
    \end{subfigure}
    \caption{Post execution decay of metaorder impact for GFI (a) and GRT (b) when using stock year configurations that minimised $\mathrm{e}_M$ (when using the PSD method to estimate $\gamma$). Power-laws were used for the trader participation distributions. Only metaorders with 3 or more child orders were used to create these plots. The dynamic impact is plotted as a function of $z$, rescaled time. The fitted curves are plotted in red. See tables \ref{tab:optimal configs GFI (metaorder) (PSD) (power)} and \ref{tab:optimal configs GRT (metaorder) (PSD) (power)} for the exact configurations used for each stock year for GFI and GRT respectively.  See table \ref{tab:calibrated impact decays (metaorder)} for the fitted values for the fitted curves of GFI and GRT respectively.}
    \label{fig:calibrated impact decays (metaorder)}
\end{figure*} 

Figure \ref{fig:calibrated impact decays (metaorder)} shows the impact decay curves for GFI and GRT when minimising $\mathrm{e}_M$ and using the PSD method to estimate $\gamma$, with the fitted curves shown as a dashed red line. The y-axis is the dynamic impact and the x-axis is $z$ (rescaled time), check equation \ref{eq:impact decay (scaled)}. Only metaorders with 3 or more child orders were included

The plots were created by binning the observations into 100 evenly spaced bins according to their value of $z$ and using the average dynamic impact of the bin as the representative value. The bin centres and their representative points are then plotted. From the Figure, it is clear that there is very little noise and the points form a very clean and smooth curves. Table \ref{tab:calibrated impact decays (metaorder)} reports the fitted values with the 95\% confidence interval half widths for the fitted curves in figures \ref{subfig:calibrated GFI impact decay (metaorder)} and \ref{subfig:calibrated GRT impact decay (metaorder)}. 

\begin{table}[h!]
\centering
\begin{tabular}{lccc}
\toprule
Description & Ticker & $Y$ & $\beta$ \\
\midrule
\makecell{Gold Fields \\ Ltd.} & GFI & 3.193 $\pm$ 0.008 & 0.159 $\pm$ 0.002\\
\makecell{Growthpoint \\ Ltd.} & GRT & 2.215 $\pm$ 0.006 & 0.177 $\pm$ 0.003\\ 
\bottomrule
\end{tabular}
\caption{Fitted values for metaorder impact decay of GFI and GRT. These are the fitted values for figures \ref{subfig:calibrated GFI impact decay (metaorder)} and \ref{subfig:calibrated GRT impact decay (metaorder)}.}
\label{tab:calibrated impact decays (metaorder)}
\end{table}

From Table \ref{tab:calibrated impact decays (metaorder)} it can be seen that both GFI and GRT have very similar fitted values for $\beta$, $0.159 \pm 0.002$ and $0.177 \pm 0.003$ respectively. These values are extremely close to the value of 0.2 value recovered in \citep{brokmann2014slowdecayimpactequity}, which analysed real metaorders. The confidence interval half widths are also extremely narrow, indicating very precise fitted values. 

\subsubsection{$\gamma \approx \alpha - 1$ ($\mathrm{e}_M$)}\label{subsubsec:alpha gamma (metaorder)}

\begin{figure*}[h!]
    \centering
    \begin{subfigure}[h]{0.45\linewidth}
        \includegraphics[width=\textwidth]{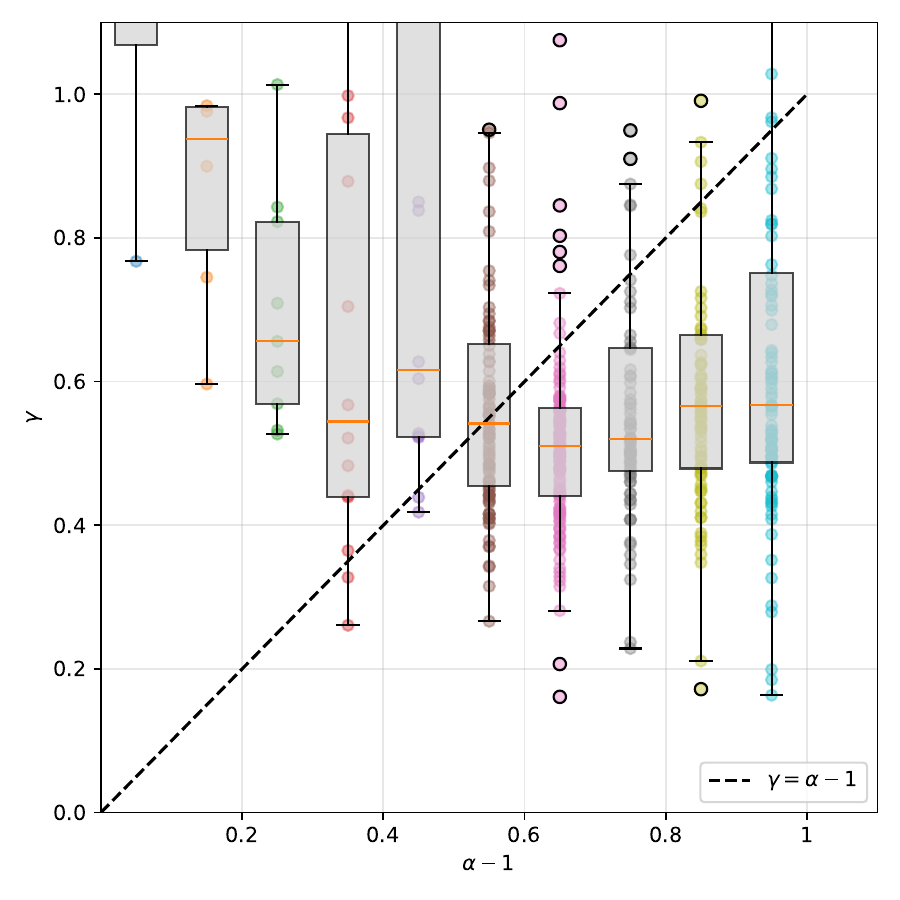}
        \caption{NLLS}
        \label{subfig:calibrated alpha gamma nlls (metaorder)}
    \end{subfigure}
    \begin{subfigure}[h]{0.45\linewidth}
        \includegraphics[width=\textwidth]{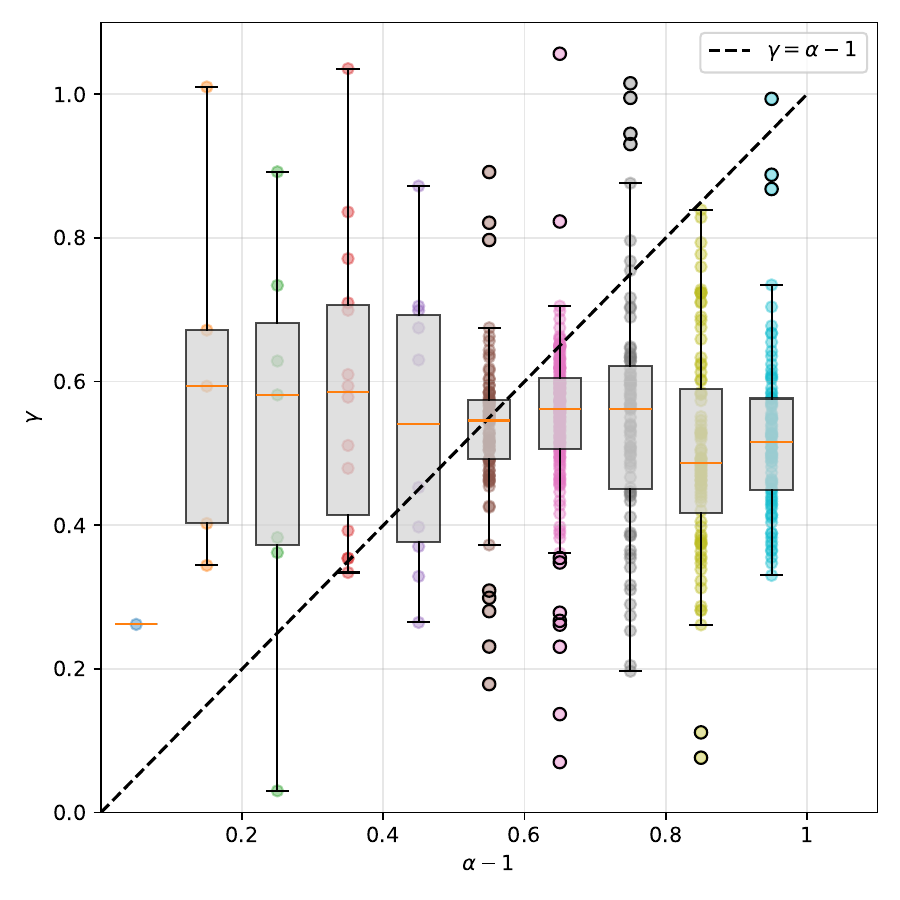}
        \caption{PSD}
        \label{subfig:calibrated alpha gamma psd (metaorder)}
    \end{subfigure}
    \caption{Scattered box plots between $\alpha -1$ and $\gamma$ when using power-law trader participation distributions, minimising $\mathrm{e}_{M}$ and estimating $\gamma$ using the NLLS (a) and PSD (b) methods respectively. Each point represents the measurement from a stock year. The median of each boxplot is shown in orange and there is a dashed black line indicating the theoretical relationship. From the plot it is clear that the LMF relation is not recovered. This result can be contrasted with the result obtained when minimising $\mathrm{e}_{LMF}$, see Figure \ref{fig:calibrated alpha gamma (LMF)}}
    \label{fig:calibrated alpha gamma (metaorder)}
\end{figure*}

Figure \ref{fig:calibrated alpha gamma (metaorder)} shows the scattered boxplots of $\gamma$ and $\alpha - 1$ when using the NLLS and PSD methods for estimating $\gamma$ with the theoretical line shown. The median of each boxpot is shown as the horizontal orange bar. The y-axis is $\gamma$ and the x-axis is $\alpha - 1$, check equation \ref{eq:alpha gamma}. 

The plots were created by first binning the observations into 10 equally spaced bins according to their $\alpha - 1$ value and then creating a box plot using the observations in each bin. From the Figure it is clear that the proposed relationship of $\gamma \approx \alpha - 1$ does not exist for either method of estimating $\gamma$. The medians are horizontal, indicating independence between $\gamma$ and $\alpha - 1$. Interestingly, the medians are all approximately equal to 0.5 which is exactly the value that we expect for most stocks \citep{SatoKanazawa2023LMFTest}.

\subsubsection{Metaorders from minimising $\mathrm{e}_M$}\label{subsubsec:e_M conclusion}

Figures \ref{fig:calibrated impact curves (metaorder)} and \ref{fig:calibrated market impact curve (metaorder)} show that the price impact of the synthetic metaorders obey the SQL. Figure \ref{fig:calibrated concave profiles (metaorder)} and Table \ref{tab:calibrated concave profiles (metaorder)} show that the synthetic metaorders have execution profiles that get close to the theoretical. Figure \ref{fig:calibrated impact decays (metaorder)} and Table \ref{tab:calibrated impact decays (metaorder)} show that the impact decay of the synthetic metaorders obey the proposed decay function \ref{eq:impact decay} and exhibit a rate of decay that is similar to the rate of 0.2 found in \cite{brokmann2014slowdecayimpactequity}. These figures show that when selecting configurations that minimise $\mathrm{e}_M$, the resulting synthetic metaorders exhibit all the stylised facts of real metaorders. However, Figure \ref{fig:calibrated alpha gamma (metaorder)} shows that under these conditions, the synthetic metaorders fail to obey the $\gamma \approx \alpha - 1$ relation of the LMF theory.

\subsection{Minimising $\mathrm{e}_{LMF}$}\label{subsec:minimising e_LMF}

\begin{table*}[t]
\centering
\begin{tabular}{lcccccccccc}
\toprule
ticker & year & N & $\delta$ & $\alpha$ & $\gamma$ & \makecell{$Q/V_D$ \\ exponent} & \makecell{$\phi$ \\ exponent} & $\beta$ & $\mathrm{e}_{M}$ (PSD) & $\mathrm{e}_{LMF}$ (PSD) \\
\midrule
GFI & 2023 & 30 & 4.00 & 1.58 & 0.57 & 0.09 & 0.97 & 0.17 & 0.66 & 0.00 \\
GFI & 2024 & 30 & 4.00 & 1.57 & 0.56 & 0.11 & 0.96 & 0.14 & 0.69 & 0.00 \\
GFI & 2025 & 50 & 4.00 & 1.65 & 0.64 & 0.02 & 0.88 & 0.15 & 0.63 & 0.00 \\
\bottomrule
\end{tabular}
\caption{Best configurations for GFI when selecting for optimal LMF results (minimising $\mathrm{e}_{LMF}$), using power-laws for the trader participation distributions and using the PSD method to estimate $\gamma$.}
\label{tab:optimal configs GFI (LMF) (PSD) (power)}
\end{table*}

\begin{table*}[t]
\centering
\begin{tabular}{lrrrrrrrrrr}
\toprule
ticker & year & N & $\delta$ & $\alpha$ & $\gamma$ & \makecell{$Q/V_D$ \\ exponent} & \makecell{$\phi$ \\ exponent} & $\beta$ & $\mathrm{e}_{M}$ (PSD) & $\mathrm{e}_{LMF}$ (PSD) \\
\midrule
GRT & 2023 & 50 & 3.00 & 1.53 & 0.48 & 0.07 & 0.80 & 0.19 & 0.58 & 0.03 \\
GRT & 2024 & 100 & 3.00 & 1.55 & 0.43 & 0.10 & 0.66 & 0.12 & 0.56 & 0.08 \\
GRT & 2025 & 500 & 2.00 & 1.52 & 0.49 & 0.12 & 0.73 & 0.13 & 0.57 & 0.02 \\
\bottomrule
\end{tabular}
\caption{Best configurations for GRT when selecting for optimal LMF results (minimising $\mathrm{e}_{LMF}$), using power-laws for the trader participation distributions and using the PSD method to estimate $\gamma$.}
\label{tab:optimal configs GRT (LMF) (PSD) (power)}
\end{table*}

Table \ref{tab:optimal configs GFI (LMF) (PSD) (power)} shows the optimal configurations for each year for GFI when minimising $\mathrm{e}_{LMF}$. From the values for $\mathrm{e}_{LMF}$ in the table, it is clear that we recovered an exact relationship of $\gamma = \alpha -1$. The error is 0 for 2023, 2024 and 2025. We get a slightly larger $\mathrm{e}_{M}$ than in Table \ref{tab:optimal configs GRT (metaorder) (PSD) (power)} for all three years. Table \ref{tab:optimal configs GRT (LMF) (PSD) (power)} shows the optimal configurations for each year for GRT when minimising $\mathrm{e}_{LMF}$. Once again, the values for $\mathrm{e}_{LMF}$ are small and represent a near perfect fit. The values for $\mathrm{e}_{M}$ are slightly larger than the ones in Table \ref{tab:optimal configs GRT (metaorder) (PSD) (power)}.

\subsubsection{SQL ($\mathrm{e}_{LMF}$)}\label{subsubsec:SQL (LMF)}

\begin{figure*}[h!]
    \centering
    \begin{subfigure}[h]{0.45\linewidth}
        \includegraphics[width=\textwidth]{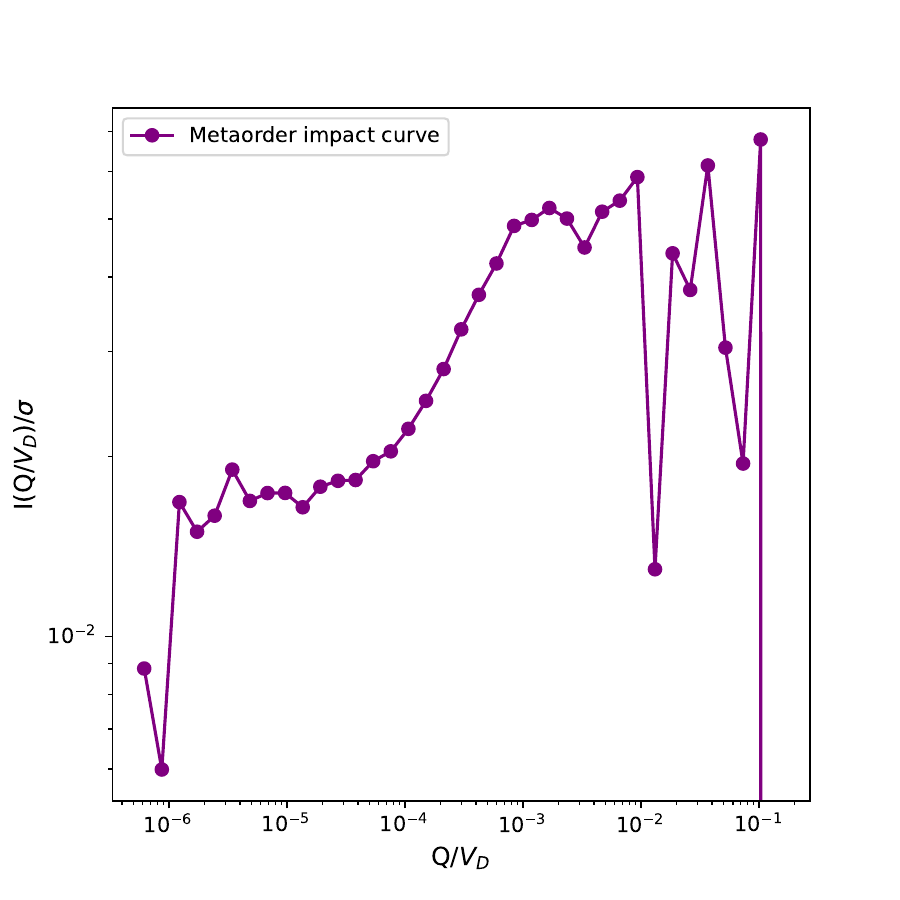}
        \caption{GFI}
        \label{subfig:calibrated GFI impact curve (LMF)}
    \end{subfigure}
    \begin{subfigure}[h]{0.45\linewidth}
        \includegraphics[width=\textwidth]{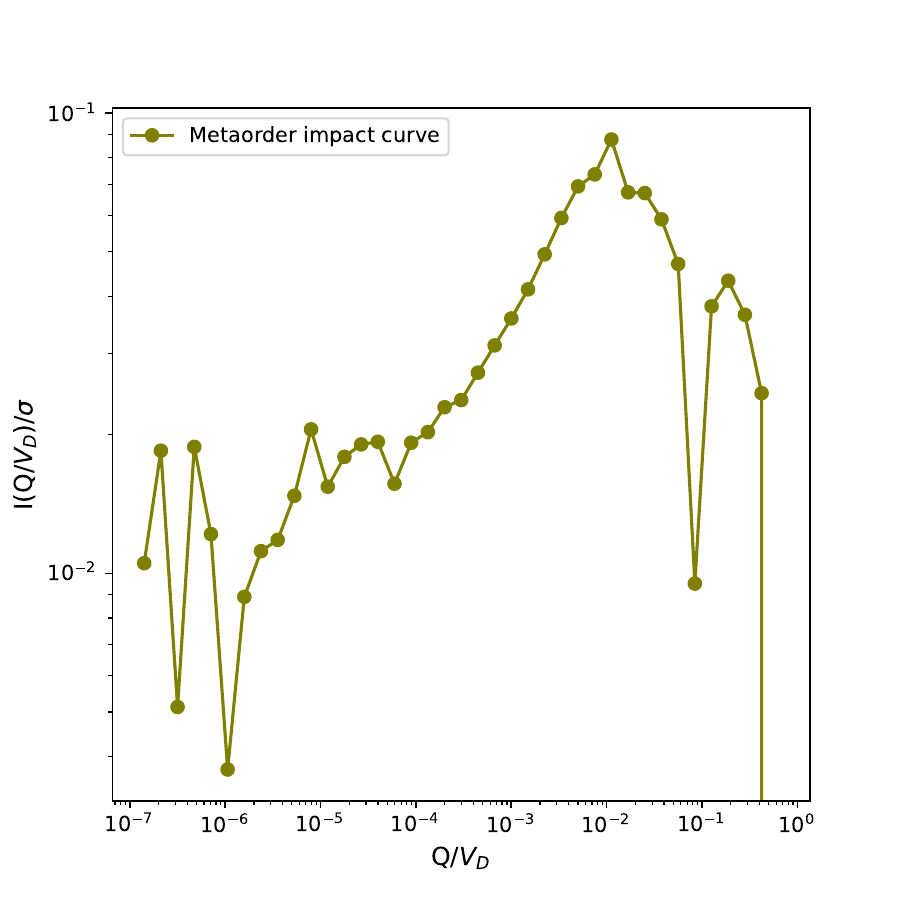}
        \caption{GRT}
        \label{subfig:calibrated GRT impact curve (LMF)}
    \end{subfigure}
    \caption{Metaorder impact curves for GFI (a) and GRT (b) when using stock year configurations that minimised $\mathrm{e}_{LMF}$ (when using the PSD method to estimate $\gamma$). Power-laws were used for the trader participation distributions. See tables \ref{tab:optimal configs GFI (LMF) (PSD) (power)} and \ref{tab:optimal configs GRT (LMF) (PSD) (power)} for the exact configurations used for each stock year for GFI and GRT respectively. Once again, the impact curves for GFI (a) and GRT (b) have noisy tails while the pooled and aggregated impact curve (Figure \ref{fig:calibrated market impact curve (LMF)}) follows the SQL closely.}
    \label{fig:calibrated impact curves (LMF)}
\end{figure*} 

Figure \ref{fig:calibrated impact curves (LMF)} shows the impact curves for GFI and GRT when minimising $\mathrm{e}_{LMF}$ and using the PSD method to estimate $\gamma$. The axes, plotting scale and plotting technique are identical to those in Section \ref{subsubsec:SQL (metaorder)}. Similarly to Figure \ref{fig:calibrated impact curves (metaorder)}, we once again see that for both large and small $Q/V_D$ there is significant noise. There is however more noise in Figure \ref{fig:calibrated impact curves (LMF)} compared to Figure \ref{fig:calibrated impact curves (metaorder)}.

\begin{figure}[h!]
    \centering
    \includegraphics[width=0.5\textwidth]{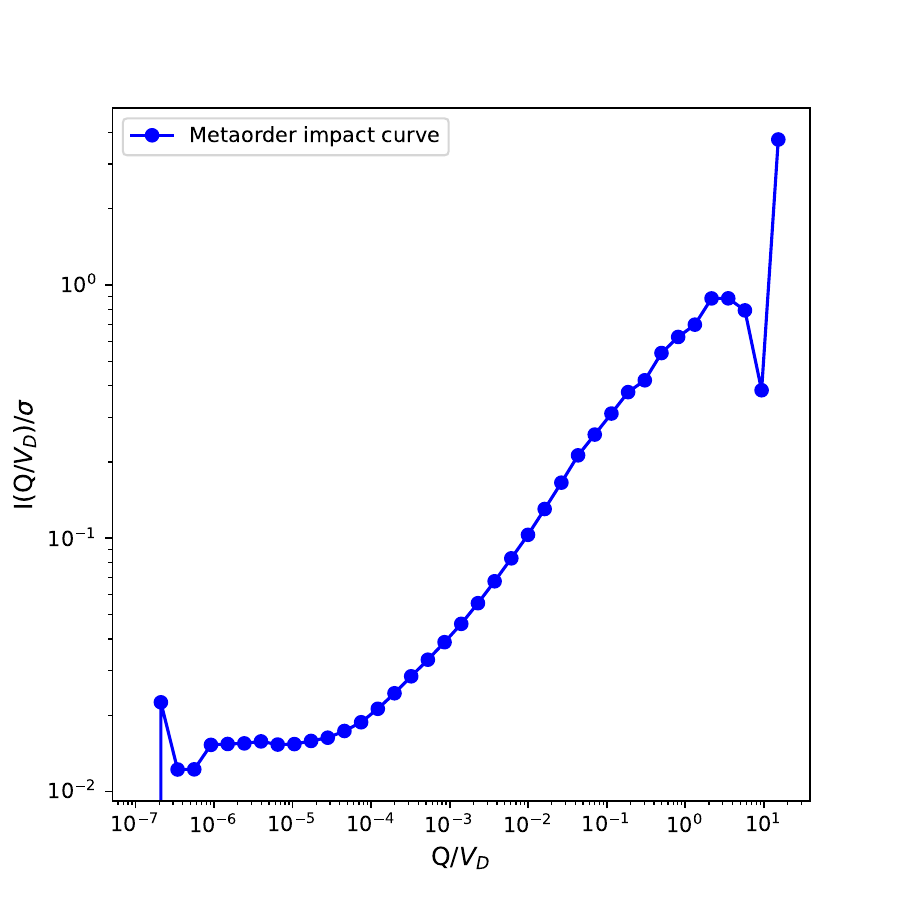}
    \caption{Metaorder impact curve for our entire universe of stocks, pooled and aggregated into a single curve, when using stock year configurations that minimised $\mathrm{e}_{LMF}$ (when using the PSD method to estimate $\gamma$). Power-laws were used for the trader participation distributions. Compared to Figure \ref{fig:calibrated impact curves (LMF)}, there is less noise in the tails and better obedience to the SQL. Similarly to Figure \ref{fig:calibrated market impact curve (metaorder)} and to the results obtained in \citep{Harvey_2017}, we see that for small relative metaorders ($Q/V_D \leq 10^{-2}$), there is a plateau where the impact is larger than the SQL suggests it should be.}
    \label{fig:calibrated market impact curve (LMF)}
\end{figure} 

Figure \ref{fig:calibrated market impact curve (LMF)} shows the metaorder impact curve for the entire universe of stocks, aggregated into a single curve. Metaorders were generated for each stock year by using the configuration that minimised $\mathrm{e}_{LMF}$ (when using the PSD method to estimate $\gamma$) for that stock year. The plots were created in the same way as Figure \ref{fig:calibrated market impact curve (LMF)}. 

\begin{table*}[h!]
\centering
\begin{tabular}{cccccc}
\toprule
\makecell{trade \\sign} & count & \makecell{proportion of \\data} & \makecell{average \\ $Q/V_D$  \\ $[\times 10^{-3}]$} & \makecell{proportion of metaorders \\ with $Q/V_D \leq 10^{-2}$} & \makecell{average $Q/V_D$ for metaorders \\ with $Q/V_D \leq 10^{-2}$  \\ $[\times 10^{-4}]$} \\
\midrule
1 & 25529711 & 50.03 & 1.342 & 39.12 & 3.55 \\
-1 & 25497441 & 49.97 & 1.341 & 39.03 & 3.54 \\
\bottomrule
\end{tabular}
\caption{Checking for trade sign imbalance in synthetic metaorders when using configurations that minimised $\mathrm{e}_{LMF}$. The data is split approximately equally between buy and sell metaorders, trade signs of 1 and -1 respectively. The average relative size of the metaorders, $Q/V_D$ is nearly identical (1.342 $\times 10^{-3}$ and $1.341 \times 10^{-3}$) between buy and sell metaorders. So it is clear that the data is not imbalanced and there is no meaningful difference in the relative sizes of the buy and sell metaorders respectively. From Figure \ref{fig:calibrated market impact curve (LMF)}, the plateau appears for $Q/V_D \leq 10^{-2}$ but it is clear that the proportion of metaorders with $Q/V_D \leq 10^{-2}$ is approximately the same for buy and sell metaorders. Additionally, the average $Q/V_D$ for that proportion is also nearly identical ($3.55 \times 10^{-4}$ and $3.54 \times 10^{-4}$) for buy and sell metaorders. These results show that imbalance in the dataset is not causing the plateau in Figure \ref{fig:calibrated market impact curve (LMF)}.}
\label{tab:trade sign imbalance table (LMF)}
\end{table*}

Figure \ref{fig:calibrated market impact curve (LMF)} is very similar to Figure \ref{fig:calibrated market impact curve (metaorder)}. It is clear that the metaorder impact obeys the SQL for $Q/V_D \geq 10^{-2}$. It is clear that for $Q/V_D < 10^{-2}$ the metaorder impact flattens significantly and forms a plateau, consistent with Figure \ref{fig:calibrated market impact curve (metaorder)}. Table \ref{tab:trade sign imbalance table (LMF)} once again compares the number of meta\-orders, the porportion of data, the average $Q/V_D$, proportion of metaorders with $Q/V_D \leq 10^{-2}$ and the average $Q/V_D$ for metaorders with $Q/V_D \leq 10^{-2}$ for buy and sell synthetic metaorders. The metaorders were constructed using configurations that minimised $\mathrm{e}_{LMF}$. From the table, the buy and sell metaorders are almost perfectly balanced (50.03\% and 49.97\% respectively), while the average relative metaorder sizes $Q/V_D$ were nearly identical ($1.342 \times 10^{-3}$ and $1.341 \times 10^{-3}$ respectively). Table \ref{tab:trade sign imbalance table (metaorder)} also shows that there is no imbalance in the proportion of buy and sell metaorders that have relative sizes $Q/V_D \leq 10^{-2}$ (39.12\% and 39.03\% respectively) and the average relative size $Q/V_D$ of this proportion is also nearly identical ($3.55 \times 10^{-4}$ and $3.54 \times 10^{-4}$) for buy and sell metaorders respectively. Similarly to Table \ref{tab:trade sign imbalance table (metaorder)}, the results in Table \ref{tab:trade sign imbalance table (LMF)} show that the plateau is unlikely to be caused by a trade sign imbalance in the synthetic metaorders. The observed plateau is however consistent with the results in Figure \ref{fig:calibrated market impact curve (metaorder)}, which used metaorders that were constructed by using configurations that minimised $\mathrm{e}_{M}$, and the results found in \citep{Harvey_2017}.

\subsubsection{Time independence ($\mathrm{e}_{LMF}$)}\label{subsubsec:time independence (LMF)}

\begin{figure*}[h!]
    \centering
    \begin{subfigure}[h]{0.45\linewidth}
        \includegraphics[width=\textwidth]{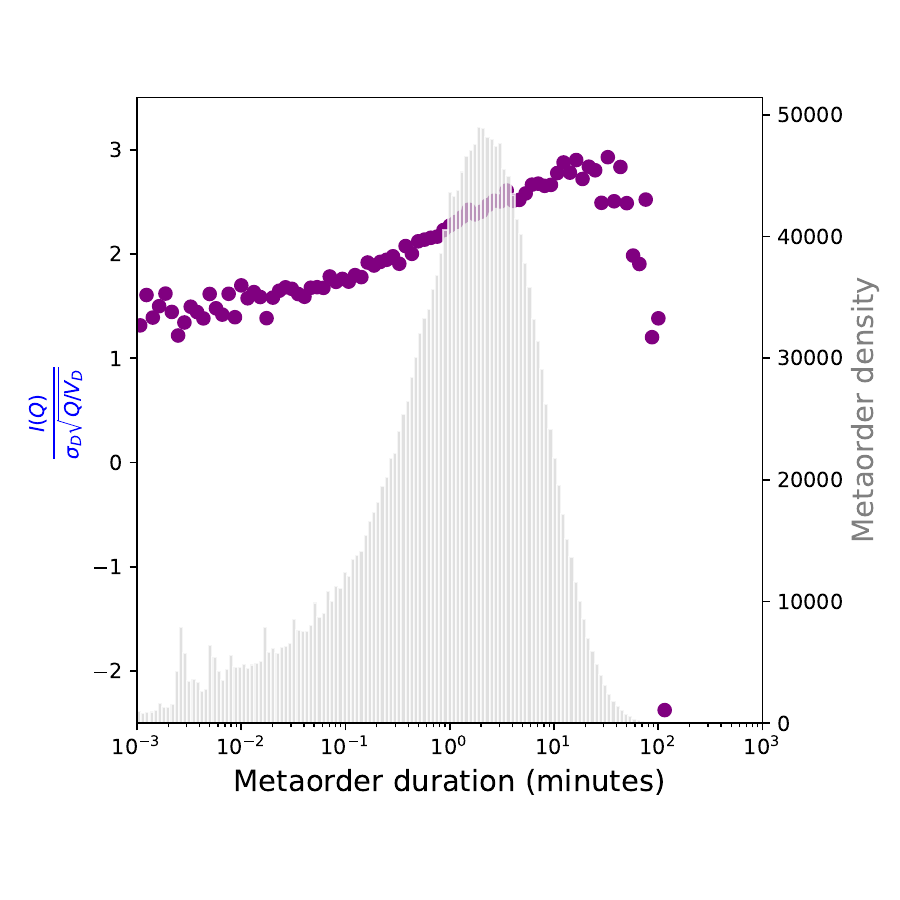}
        \caption{GFI}
        \label{subfig:calibrated GFI time independence (LMF)}
    \end{subfigure}
    \begin{subfigure}[h]{0.45\linewidth}
        \includegraphics[width=\textwidth]{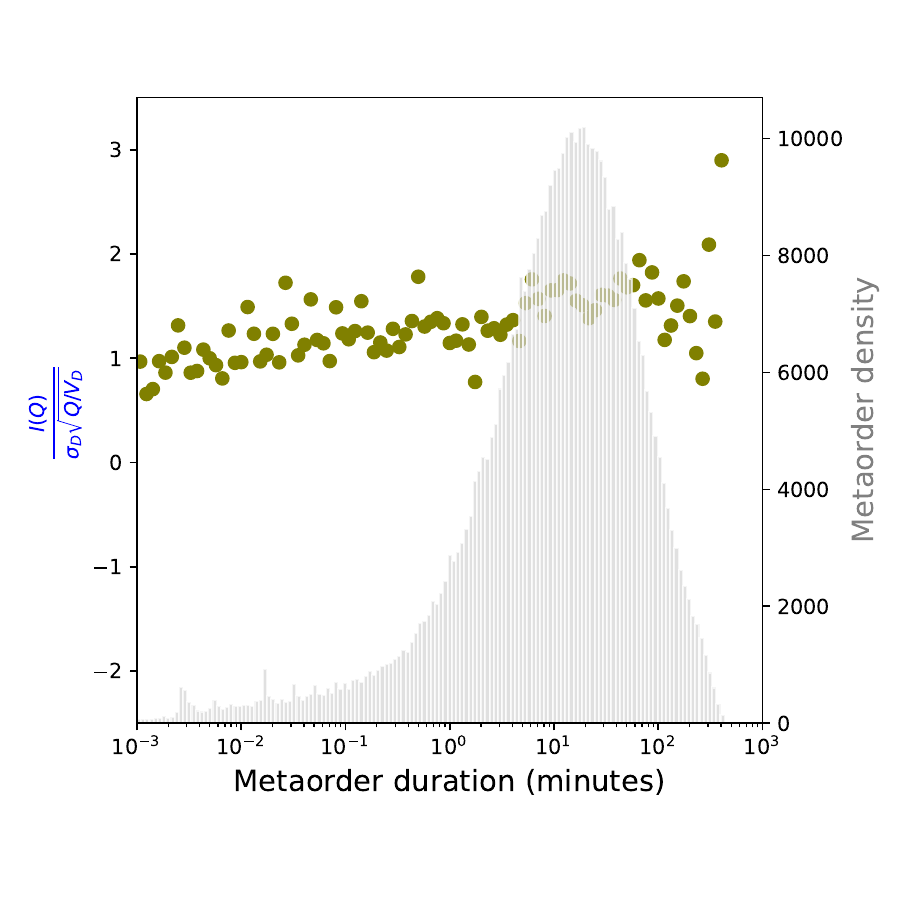}
        \caption{GRT}
        \label{subfig:calibrated GRT time independence (LMF)}
    \end{subfigure}
    \caption{Time independence of metaorder impact plots for GFI (a) and GRT (b) when using stock year configurations that minimised $\mathrm{e}_{LMF}$ (when using the PSD method to estimate $\gamma$). Power-laws were used for the trader participation distributions. The average metaorder duration for GFI was approximately $10^{0.5}$ (3) minutes and for GRT it is approximately $10^{1.5}$ (32) minutes. While the average metaorder duration for GFI is apprximately equal, for GRT, stock year configuration that minimise $\mathrm{e}_{LMF}$ clearly result in metaorders with shorter average durations than in Figure \ref{fig:calibrated time independence (metaorder)}.} 
    \label{fig:calibrated time independence (LMF)}
\end{figure*} 

Figure \ref{fig:calibrated time independence (metaorder)} shows the independence of metaorder impact with respect to the metaorder duration for GFI and GRT when using metaorder configurations that minimise $\mathrm{e}_{LMF}$, power-law trader participation distributions and the PSD method to estimate $\gamma$. The axes, plotting scale and plotting technique is identical to those in Section \ref{subsubsec:time independence (metaorder)}.

From Figure \ref{fig:calibrated time independence (metaorder)}, it is clear that the relationship between the dynamic impact and the metaorder duration is approximately flat and indicates independence, similarly to Figure \ref{fig:calibrated time independence (metaorder)} and to the results obtained in \citep{Maitrieretal2025}. It is also clear that the average metaorder duration for GFI (Figure \ref{subfig:calibrated GFI time independence (LMF)}) is approximately 3 minutes, and for GRT (Figure \ref{subfig:calibrated GRT time independence (LMF)}) it is approximately 32 minutes. GFI has very similar average durations but in GRT the average duration is clearly smaller in Figure \ref{fig:calibrated time independence (LMF)} compared to Figure \ref{fig:calibrated time independence (metaorder)} . Both sub-figures show long left tails with short right tails, indicating far more metaorders with short durations compared to long durations. This tail behaviour similar to that in Figure \ref{fig:calibrated time independence (metaorder)}. 

\subsubsection{Execution profile ($\mathrm{e}_{LMF}$)}\label{subsubsec:execution profile (LMF)}

\begin{figure*}[h!]
    \centering
    \begin{subfigure}[h]{0.45\linewidth}
        \includegraphics[width=\textwidth]{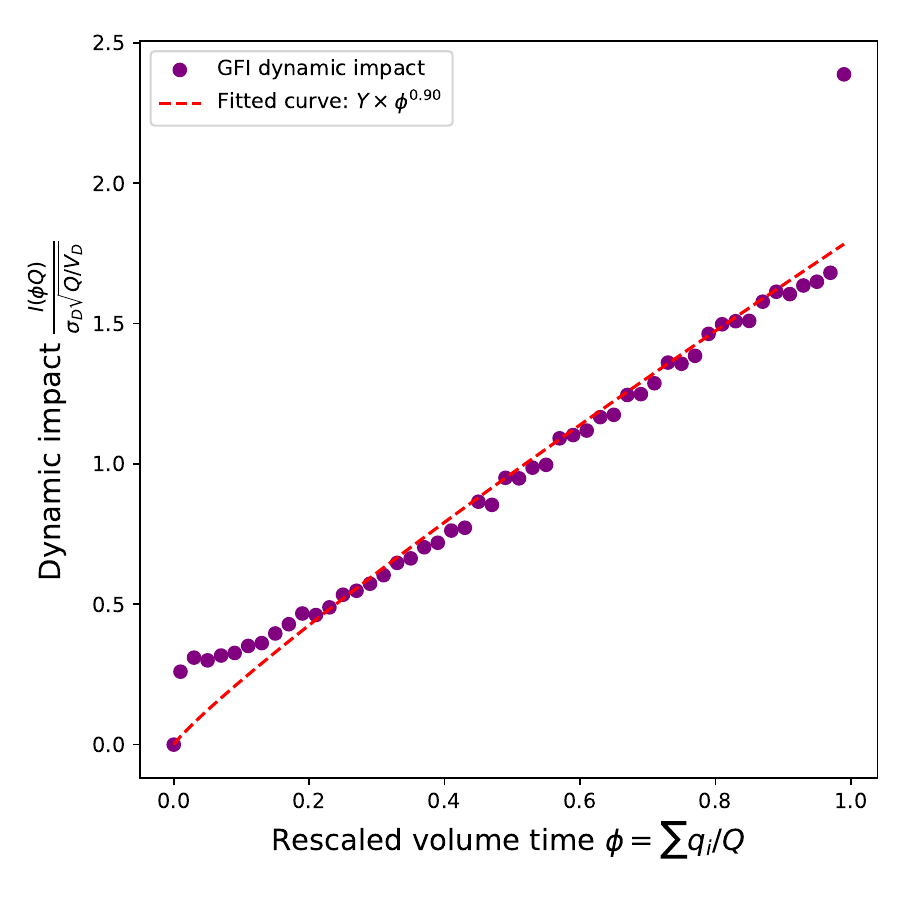}
        \caption{GFI}
        \label{subfig:calibrated GFI concave profile (LMF)}
    \end{subfigure}
    \begin{subfigure}[h]{0.45\linewidth}
        \includegraphics[width=\textwidth]{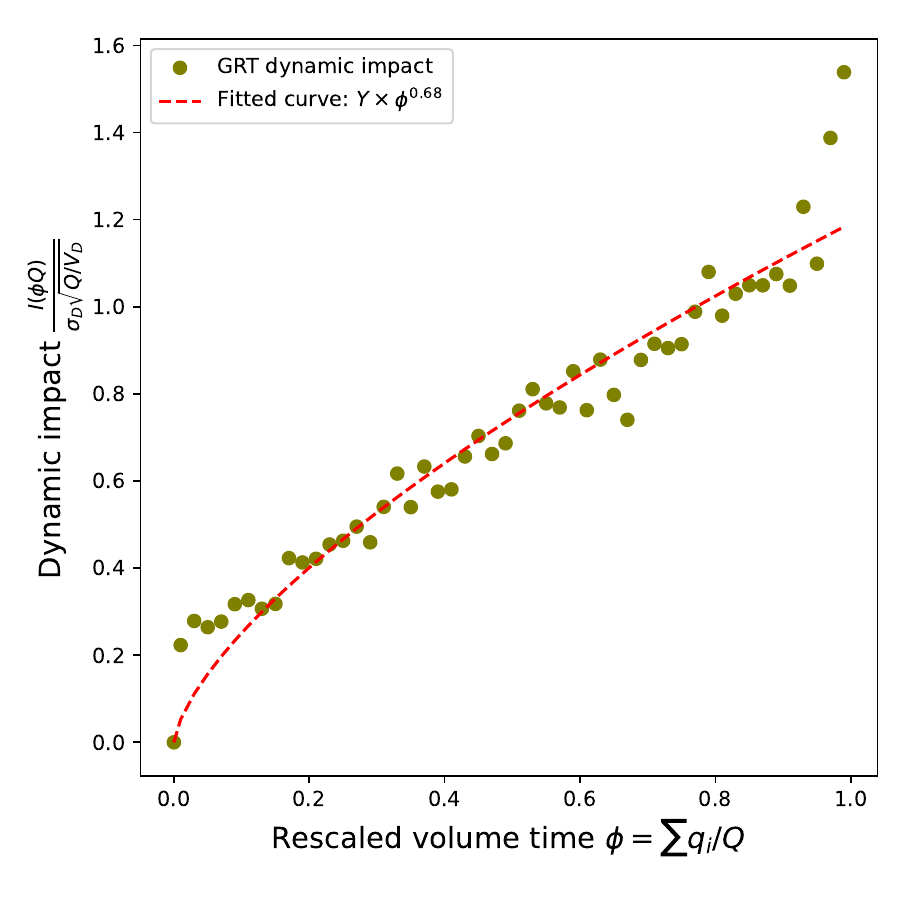}
        \caption{GRT}
        \label{subfig:calibrated GRT concave profile (LMF)}
    \end{subfigure}
    \caption{Concave profile of metaorder impact for GFI (a) and GRT (b) when using stock year configurations that minimised $\mathrm{e}_{LMF}$ (when using the PSD method to estimate $\gamma$). Power-laws were used for the trader participation distributions. Only metaorders with 3 or more child orders were used to create these plots. The dynamic impact is plotted as a function of $\phi$, the proportion of the metaorder which has been executed. The fitted curves are also plotted in red. See tables \ref{tab:optimal configs GFI (LMF) (PSD) (power)} and \ref{tab:optimal configs GRT (LMF) (PSD) (power)} for the exact configurations used for each stock year for GFI and GRT respectively. See Table \ref{tab:calibrated concave profiles (LMF)} for the fitted values for the fited curves of GFI and GRT respectively.}
    \label{fig:calibrated concave profiles (LMF)}
\end{figure*} 

Figure \ref{fig:calibrated concave profiles (LMF)} shows the execution profiles for GFI and GRT when minimising $\mathrm{e}_{LMF}$ and using the PSD method to estimate $\gamma$, with the fitted curves shown as a dashed red line. The axes, plotting scale and plotting technique are identical to those in Section \ref{subsubsec:execution profile (metaorder)}. Once again, only metaorders with 3 or more child orders were included. 

It is clear that there is more noise in the points for GRT (Figure \ref{subfig:calibrated GRT concave profile (LMF)}) than for GFI (Figure \ref{subfig:calibrated GFI concave profile (LMF)}). This may however be due there being too few observations per bin as a result of there being much fewer metaorders for GRT compared to GFI. Despite this however, the fits in both cases are good and there are no clear deviations from the points. Table \ref{tab:calibrated concave profiles (LMF)} reports the fitted values with the 95\% confidence interval half widths for the fitted curves in figures \ref{subfig:calibrated GFI concave profile (LMF)} and \ref{subfig:calibrated GRT concave profile (LMF)}. 

\begin{table}[h!]
\centering
\begin{tabular}{lccc}
\toprule
Description & Ticker & $Y$ & \makecell{$\phi$ \\ exponent} \\
\midrule
\makecell{Gold Fields \\ Ltd.} & GFI & 1.799 $\pm$ 0.038 & 0.896 $\pm$ 0.042\\
\makecell{Growthpoint \\ Ltd.} & GRT & 1.192 $\pm$ 0.027 & 0.677 $\pm$ 0.038\\ 
\bottomrule
\end{tabular}
\caption{Fitted values for metaorder execution profiles of GFI and GRT. These are the fitted values for figures \ref{subfig:calibrated GFI concave profile (LMF)} and \ref{subfig:calibrated GRT concave profile (LMF)}.}
\label{tab:calibrated concave profiles (LMF)}
\end{table}

From the Table it is clear that the exponent of $\phi$ for GFI is $0.896 \pm 0.042$ and for GRT the exponent is $0.677 \pm 0.038$ while the exponents when selecting configurations that minimised $\mathrm{e}_M$ we obtained values of $0.802 \pm 0.044$ and $0.704 \pm 0.046$ respectively. It is therefore clear that when compared to configurations selected to minimise $\mathrm{e}_M$, the exponent of $\phi$ is further from the theoretical value of 0.5 for GFI, while for GRT it is very slightly closer. Despite this, these values still indicate that there is concavity in the execution profiles. In both cases, the confidence interval half widths are narrow which again indicates that the fitted values are precise. 

\subsubsection{Impact decay ($\mathrm{e}_{LMF}$)}\label{subsubsec:impact decay (LMF)}

\begin{figure*}[h!]
    \centering
    \begin{subfigure}[h]{0.45\linewidth}
        \includegraphics[width=\textwidth]{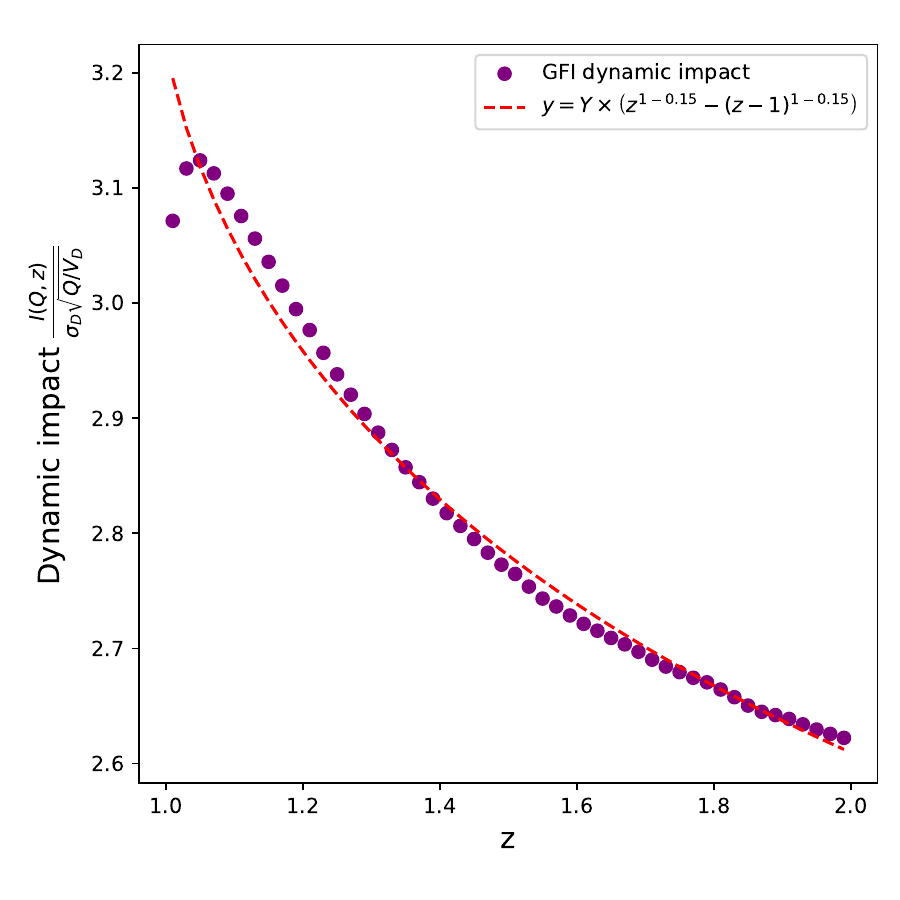}
        \caption{GFI}
        \label{subfig:calibrated GFI impact decay (LMF)}
    \end{subfigure}
    \begin{subfigure}[h]{0.45\linewidth}
        \includegraphics[width=\textwidth]{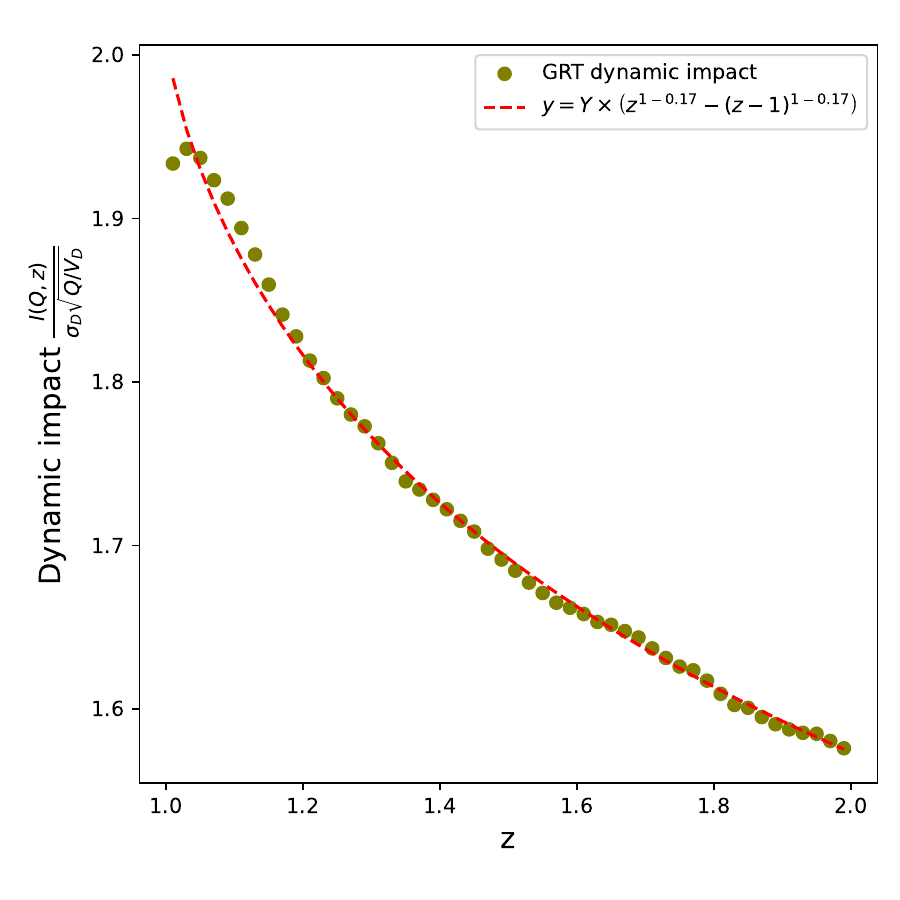}
        \caption{GRT}
        \label{subfig:calibrated GRT impact decay (LMF)}
    \end{subfigure}
    \caption{Post execution decay of metaorder impact for GFI (a) and GRT (b) when  using stock year configurations that minimised $\mathrm{e}_{LMF}$ (when using the PSD method to estimate $\gamma$). Power-laws were used for the trader participation distributions.  Only metaorders with 3 or more child orders were used to create these plots. The dynamic impact is plotted as a function of $z$, rescaled time. The fitted curves are plotted in red. See tables \ref{tab:optimal configs GFI (LMF) (PSD) (power)} and \ref{tab:optimal configs GRT (LMF) (PSD) (power)} for the exact configurations used for each stock year for GFI and GRT respectively.  See table \ref{tab:calibrated impact decays (LMF)} for the fitted values for the fitted curves of GFI and GRT respectively.}
    \label{fig:calibrated impact decays (LMF)}
\end{figure*} 

Figure \ref{fig:calibrated impact decays (LMF)} shows the impact decay curves for GFI and GRT when minimising $\mathrm{e}_{LMF}$ and using the PSD method to estimate $\gamma$, with the fitted curves shown as a dashed red line. The axes, plotting scale and plotting technique are identical to those in Section \ref{subsubsec:impact decay (metaorder)}. Once again, only metaorders with 3 or more child orders were included. 

When compared to Figure \ref{fig:calibrated impact decays (metaorder)}, Figure \ref{fig:calibrated impact decays (LMF)} has less deviation in the points from the fitted curve, particularly for GRT. Despite this, the points still form a smooth curve. Table \ref{tab:calibrated impact decays (metaorder)} reports the fitted values with the 95\% confidence interval half widths for the fitted curves in figures \ref{subfig:calibrated GFI impact decay (LMF)} and \ref{subfig:calibrated GRT impact decay (LMF)}. 

\begin{table}[h!]
\centering
\begin{tabular}{lccc}
\toprule
Description & Ticker & $Y$ & $\beta$ \\
\midrule
\makecell{Gold Fields \\ Ltd.} & GFI & 3.231 $\pm$ 0.010 & 0.146 $\pm$ 0.003\\
\makecell{Growthpoint \\ Ltd.} & GRT & 2.012 $\pm$ 0.004 & 0.167 $\pm$ 0.002\\ 
\bottomrule
\end{tabular}
\caption{Fitted values for metaorder impact decay of GFI and GRT. These are the fitted values for figures \ref{subfig:calibrated GFI impact decay (LMF)} and \ref{subfig:calibrated GRT impact decay (LMF)}.}
\label{tab:calibrated impact decays (LMF)}
\end{table}

From Table \ref{tab:calibrated impact decays (LMF)}, the specific fitted values for $\beta$ are $0.146 \pm 0.003$ and $0.167 \pm 0.002$ for GFI and GRT respectively. These values are smaller than the fitted values of $0.159 \pm 0.002$ and $0.177 \pm 0.003$ (Table \ref{tab:calibrated impact decays (metaorder)}) which were obtained when minimising for $\mathrm{e}_M$ and using the PSD method to estimate $\gamma$. This indicates a more linear decay. 

\subsubsection{$\gamma \approx \alpha - 1$ ($\mathrm{e}_{LMF}$)}\label{subsubsec:alpha gamma (LMF)}

\begin{figure*}[h!]
    \centering
    \begin{subfigure}[h]{0.45\linewidth}
        \includegraphics[width=\textwidth]{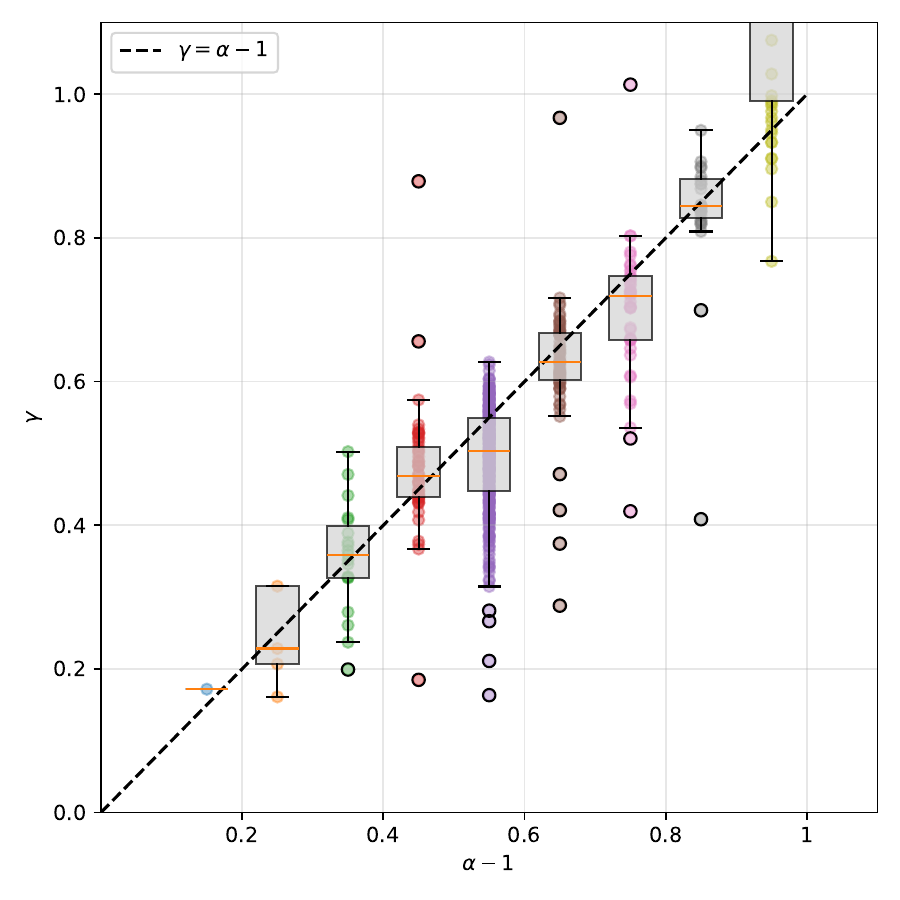}
        \caption{NLLS}
        \label{subfig:calibrated alpha gamma nlls (LMF)}
    \end{subfigure}
    \begin{subfigure}[h]{0.45\linewidth}
        \includegraphics[width=\textwidth]{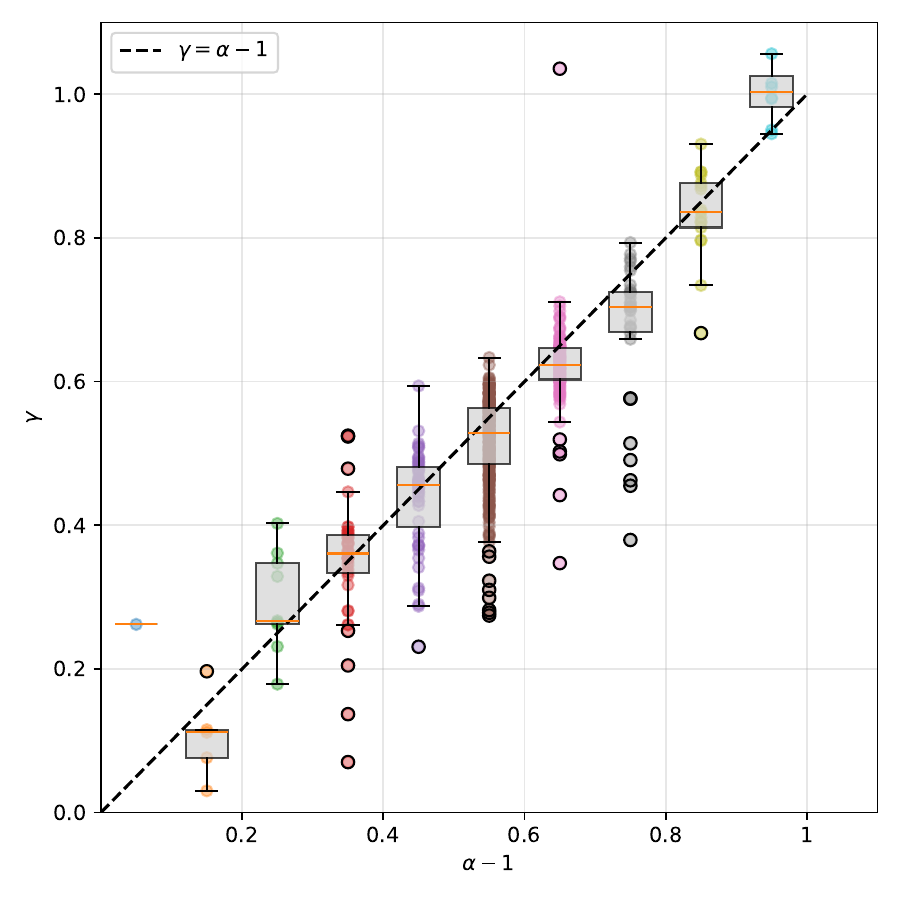}
        \caption{PSD}
        \label{subfig:calibrated alpha gamma psd (LMF)}
    \end{subfigure}
    \caption{Scattered box plots between $\alpha -1$ and $\gamma$ when using power-law trader participation distributions, minimising $\mathrm{e}_{LMF}$ and estimating $\gamma$ using the NLLS (a) and PSD (b) methods respectively. The median of each boxplot is shown in orange and there is a dashed black line indicating the theoretical relationship. It is clear that the LMF relation is recovered, the medians of each boxplot are almost perfectly aligned with the theoretical. While the fact we get a better fit than in Figure \ref{fig:calibrated alpha gamma (metaorder)} should not be a surprise, the fact that we obtained an almost perfect relation is a surprise. Each point is the measurement from a stock year, so it is clear that we do not recover a perfect $\gamma = \alpha - 1$ relation. But despite this, the aggregate market effect is an almost perfect relation.}
    \label{fig:calibrated alpha gamma (LMF)}
\end{figure*}

Figure \ref{fig:calibrated alpha gamma (LMF)} shows the scattered boxplots of $\gamma$ and $\alpha - 1$ when using the NLLS and PSD methods for estimating $\gamma$ with the theoretical line shown. Each point represents the measurement from a stock year. The median of each boxpot is shown as the horizontal orange bar. The y-axis is $\gamma$ and the x-axis is $\alpha - 1$, check equation \ref{eq:alpha gamma}. 

The plots were created in the same way as for Figure \ref{fig:calibrated alpha gamma (metaorder)}. From the Figure it is clear that unlike when minimising $\mathrm{e}_M$, the proposed relationship of $\gamma \approx \alpha - 1$ holds for both methods of estimating $\gamma$. Although, this result is tautological since we specifically chose configurations that would result in the best $\gamma \approx \alpha - 1$ relation. It is worth noting that while each stock year does not have a perfect $\gamma = \alpha - 1$ relation, it is clear that the aggregate effect is an almost perfect relation. However, this may also be a sign of possible overfitting.   

\subsubsection{Metaorders from minimising $\mathrm{e}_{LMF}$}\label{subsubsec:e_LMF conclusion}

Figures \ref{fig:calibrated impact curves (LMF)} and \ref{fig:calibrated market impact curve (LMF)} show that the LMF-selected configurations retain a pooled square-root-like impact relation, although the stock-level curves remain heterogeneous. Figure \ref{fig:calibrated concave profiles (LMF)} and Table \ref{tab:calibrated concave profiles (LMF)} show concave but more nearly linear execution profiles than under $\mathrm{e}_M$. Figure \ref{fig:calibrated impact decays (LMF)} and Table \ref{tab:calibrated impact decays (LMF)} show decay with visible departures from the fitted form. Figure \ref{fig:calibrated alpha gamma (LMF)} shows close agreement with $\gamma \approx \alpha-1$, but this is a targeted in-sample compatibility result because the same discrepancy selects the configurations.

\subsection{Discussion}\label{subsec:Discussion}

The two selection criteria answer different questions. Minimising $\mathrm{e}_M$ selects configurations that agree most closely with the chosen impact targets, but these configurations do not recover $\gamma \approx \alpha-1$. Minimising $\mathrm{e}_{LMF}$ selects configurations that agree with the LMF exponent relation by construction, while retaining a pooled square-root-like impact relation and concave execution profiles, although the stock-level execution and decay diagnostics are weaker. Neither objective independently identifies the synthetic partitions as the latent metaorders.

\section{Conclusion}\label{sec:conclusion}

In Section \ref{sec:synthetic} we have shown that we are able to generate synthetic metaorders from public TAQ data using the methods proposed by \citet{Maitrieretal2025}. In Section \ref{sec:calibration} we lay out the methods used to calibrate Algorithms \ref{alg:mapping} and \ref{alg:synthetic_metaorders} via a grid search over the parameters of the algorithms. In Section \ref{sec:calibration results} we discuss the results from the calibration and explain the selection procedure. 

The pooled curves in Figures \ref{fig:calibrated market impact curve (metaorder)} and \ref{fig:calibrated market impact curve (LMF)} are square-root-like over the stated volume range, while the individual-stock curves remain heterogeneous. Figures \ref{fig:calibrated time independence (metaorder)} and \ref{fig:calibrated time independence (LMF)} show no strong visible duration trend in the binned summaries. The execution profiles are concave, and the post-execution profiles decay over the observed window, although the fitted exponents and fit quality depend on the calibration objective. These results establish aggregate compatibility with several metaorder impact regularities; they do not identify the synthetic partitions as the true latent metaorders.

The impact-selected configurations do not recover $\gamma \approx \alpha-1$. The LMF-selected configurations can be made closely consistent with this relation, but the agreement is a targeted calibration result because $\mathrm{e}_{LMF}$ is used to choose the configurations. The comparison therefore supports compatibility of anonymous JSE order flow with the LMF relation within the synthetic reconstruction class, rather than an account-level validation of the microscopic mechanism.

Recovering the selected impact stylised facts alone does not guarantee an LMF-consistent exponent relation. Conversely, selecting on the LMF discrepancy produces configurations that remain compatible with several aggregate impact regularities, but this does not independently recover the latent order-splitting mechanism.

The results are consistent with related agent-based modelling work where the metaorders were modelled as learnt TWAP strategies executed by multiple interacting reinforcement learning agents \cite{Dicks2023ManyLearningAgents}, and then aggregated in a reactive agent-based model \cite{DicksPaskaramoorthyGebbie2023} (in particular their Figure 6). They demonstrate how the decay in the trade sign autocorrelations caused by metaorders mixing with other agents can be calibrated to fit the empirical data while retaining little memory in the price fluctuations, and recover a square root price impact law and hence the required concavity in the price impact of orders.

Related reaction-diffusion and latent-order-book models provide a broader modelling context for concave impact and persistent order flow \cite{DianaGebbie2025,Mastromatteo2014AgentBasedLiquidity,Donier2015FullyConsistentImpact}. These models do not validate the synthetic trader partition used here, but they show how the empirical regularities considered in this paper can arise in other microstructural constructions.

\section{Data availability}

The transaction and Level 1 quote data used in this study were accessed through the BMLL Data Lab environment \citep{bmllpython2026}. The underlying market data are not redistributed with this paper and remain subject to the data provider's access conditions. The stock-universe information and the processed calibration outputs used to support the reported analysis are documented in the accompanying code repository \citep{goliath_github_repo}.

\section{Code and reproducibility}

The Python modules and notebooks used for data processing, synthetic metaorder construction, calibration, estimation, and figure generation are available in the project repository \citep{goliath_github_repo}. The repository also contains the committed calibration-results table used to trace the reported parameter estimates. Reproducing the analysis from raw market data requires access to the BMLL environment and the corresponding data products. Because the public repository is a moving code base, an immutable release or archived snapshot should be used with the final arXiv submission.

\section{Acknowledgements}

We sincerely thank Jean-Phillipe Bouchaud for some helpful feedback. We would like to thank Nasheen Sharma and Vuyo Mashiqa from the JSE, and the JSE for supporting the project. We thank Daniela Stevenson for suggestions and feedback. 

\bibliography{Preprint-v3.3.0}

@misc{Maitrieretal2025,
      title={Generating realistic metaorders from public data}, 
      author={Guillaume Maitrier and Grégoire Loeper and Jean-Philippe Bouchaud},
      year={2025},
      eprint={2503.18199},
      archivePrefix={arXiv},
      primaryClass={q-fin.TR},
      url={https://arxiv.org/abs/2503.18199}, 
}

@article{LilloMikeFarmer2005,
  title = {Theory for long memory in supply and demand},
  author = {Lillo, Fabrizio and Mike, Szabolcs and Farmer, J. Doyne},
  journal = {Phys. Rev. E},
  volume = {71},
  issue = {6},
  pages = {066122},
  numpages = {11},
  year = {2005},
  month = {Jun},
  publisher = {American Physical Society},
  doi = {10.1103/PhysRevE.71.066122},
  url = {https://link.aps.org/doi/10.1103/PhysRevE.71.066122}
}

@article{Clausetetal2009,
   title={Power-Law Distributions in Empirical Data},
   volume={51},
   ISSN={1095-7200},
   url={http://dx.doi.org/10.1137/070710111},
   DOI={10.1137/070710111},
   number={4},
   journal={SIAM Review},
   publisher={Society for Industrial & Applied Mathematics (SIAM)},
   author={Clauset, Aaron and Shalizi, Cosma Rohilla and Newman, M. E. J.},
   year={2009},
   month=nov, pages={661--703} }

@article{anomalous_price_impact,
   title={Anomalous Price Impact and the Critical Nature of Liquidity in Financial Markets},
   volume={1},
   ISSN={2160-3308},
   url={http://dx.doi.org/10.1103/PhysRevX.1.021006},
   DOI={10.1103/physrevx.1.021006},
   number={2},
   journal={Physical Review X},
   publisher={American Physical Society (APS)},
   author={Tóth, B. and Lempérière, Y. and Deremble, C. and de Lataillade, J. and Kockelkoren, J. and Bouchaud, J.-P.},
   year={2011},
   month=oct }

@misc{2014marketimpactslifecycle,
      title={Market impacts and the life cycle of investors orders}, 
      author={Emmanuel Bacry and Adrian Iuga and Matthieu Lasnier and Charles-Albert Lehalle},
      year={2014},
      eprint={1412.0217},
      archivePrefix={arXiv},
      primaryClass={q-fin.TR},
      url={https://arxiv.org/abs/1412.0217}, 
}

@article{LilloFarmer2004,
  title        = {The Long Memory of the Efficient Market},
  author       = {Lillo, Fabrizio and Farmer, J. Doyne},
  journal      = {Studies in Nonlinear Dynamics and Econometrics},
  volume       = {8},
  number       = {3},
  year         = {2004},
  doi          = {10.2202/1558-3708.1144},
  publisher    = {De Gruyter}
}

@article{WilcoxGebbie2014,
  title        = {Hierarchical causality in financial economics},
  author       = {Wilcox, Diane and Gebbie, Tim},
  journal      = {arXiv preprint arXiv:1408.5585},
  year         = {2014},
  doi          = {10.48550/arXiv.1408.5585}
}

@article{Lillo2023Decoding,
  author       = {Lillo, Fabrizio},
  title        = {Decoding the Dynamics of Supply and Demand},
  journal      = {Physics},
  volume       = {16},
  pages        = {192},
  year         = {2023},
  publisher    = {American Physical Society},
  url          = {https://physics.aps.org/articles/v16/192}
}

@misc{zarinelli2014,
      title={Beyond the square root: Evidence for logarithmic dependence of market impact on size and participation rate}, 
      author={Elia Zarinelli and Michele Treccani and J. Doyne Farmer and Fabrizio Lillo},
      year={2014},
      eprint={1412.2152},
      archivePrefix={arXiv},
      primaryClass={q-fin.TR},
      url={https://arxiv.org/abs/1412.2152}, 
}

@article{SatoKanazawa2023,
  author       = {Sato, Yuki and Kanazawa, Kiyoshi},
  title        = {Quantitative statistical analysis of order-splitting behavior of individual trading accounts in the Japanese stock market over nine years},
  journal      = {Physical Review Research},
  volume       = {5},
  number       = {4},
  pages        = {043131},
  year         = {2023},
  doi          = {10.1103/PhysRevResearch.5.043131},
  url          = {https://link.aps.org/doi/10.1103/PhysRevResearch.5.043131},
  publisher    = {American Physical Society}
}

@article{SatoKanazawa2023LMFTest,
  author       = {Sato, Yuki and Kanazawa, Kiyoshi},
  title        = {Inferring Microscopic Financial Information from the Long Memory in Market-Order Flow: A Quantitative Test of the Lillo-Mike-Farmer Model},
  journal      = {Physical Review Letters},
  volume       = {131},
  number       = {19},
  pages        = {197401},
  year         = {2023},
  doi          = {10.1103/PhysRevLett.131.197401},
  url          = {https://link.aps.org/doi/10.1103/PhysRevLett.131.197401},
  publisher    = {American Physical Society}
}

@article{Bouchaud2025UniversalLaw,
  author       = {Bouchaud, Jean-Philippe},
  title        = {The Universal Law Behind Market Price Swings},
  journal      = {Physics},
  volume       = {18},
  pages        = {196},
  year         = {2025},
  url          = {https://physics.aps.org/articles/v18/196},
  publisher    = {American Physical Society}
}

@article{DicksPaskaramoorthyGebbie2023,
  author       = {Dicks, Matthew and Paskaramoorthy, Andrew and Gebbie, Tim},
  title        = {A simple learning agent interacting with an agent-based market model},
  journal      = {Physica A: Statistical Mechanics and its Applications},
  volume       = {633},
  pages        = {129363},
  year         = {2024},
  doi          = {10.1016/j.physa.2023.129363},
  url          = {https://doi.org/10.1016/j.physa.2023.129363},
  publisher    = {Elsevier}
}

@article{DianaGebbie2025,
title = {Non-uniformly sampled simulated price impact of an order-book},
journal = {Journal of Computational and Applied Mathematics},
volume = {456},
pages = {116202},
year = {2025},
issn = {0377-0427},
doi = {https://doi.org/10.1016/j.cam.2024.116202},
url = {https://www.sciencedirect.com/science/article/pii/S0377042724004515},
author = {Derick Diana and Tim Gebbie},
}

@manual{bmllpython2026,
  author       = {{BMLL Technologies}},
  title        = {{BMLL Data Lab Python Development Environment}},
  year         = {2026},
  url          = {https://www.bmlltech.com/},
  note         = {Accessed: 2026-02-02}
}

@article{Dicks2023ManyLearningAgents,
  title        = {Many learning agents interacting with an agent-based market model},
  author       = {Dicks, Matthew and Paskaramoorthy, Andrew and Gebbie, Tim},
  year         = {2023},
  journal      = {arXiv preprint arXiv:2303.07393},
  archivePrefix= {arXiv},
  eprint       = {2303.07393},
  primaryClass = {q-fin.TR},
  url          = {https://arxiv.org/abs/2303.07393},
  note         = {arXiv:2303.07393v4, revised 14 August 2024}
}

@article{Mastromatteo2014AgentBasedLiquidity,
  title         = {Agent-based models for latent liquidity and concave price impact},
  author        = {Mastromatteo, Iacopo and Toth, Bence and Bouchaud, Jean-Philippe},
  journal       = {Physical Review E},
  volume        = {89},
  number        = {4},
  pages         = {042805},
  year          = {2014},
  publisher     = {American Physical Society},
  doi           = {10.1103/PhysRevE.89.042805},
  url           = {https://doi.org/10.1103/PhysRevE.89.042805}
}

@article{Donier2015FullyConsistentImpact,
  title         = {A fully consistent, minimal model for non-linear market impact},
  author        = {Donier, Jonathan and Bonart, Julius and Mastromatteo, Iacopo and Bouchaud, Jean-Philippe},
  journal       = {Quantitative Finance},
  volume        = {15},
  number        = {7},
  pages         = {1109--1121},
  year          = {2015},
  publisher     = {Taylor \& Francis},
  doi           = {10.1080/14697688.2015.1032540},
  url           = {https://doi.org/10.1080/14697688.2015.1032540}
}

@misc{brokmann2014slowdecayimpactequity,
      title={Slow decay of impact in equity markets}, 
      author={X. Brokmann and E. Serie and J. Kockelkoren and J. -P. Bouchaud},
      year={2014},
      eprint={1407.3390},
      archivePrefix={arXiv},
      primaryClass={q-fin.TR},
      url={https://arxiv.org/abs/1407.3390}, 
}

@article{Harvey_2017,
   title={Deviations in expected price impact for small transaction volumes under fee restructuring},
   volume={471},
   ISSN={0378-4371},
   url={http://dx.doi.org/10.1016/j.physa.2016.11.042},
   DOI={10.1016/j.physa.2016.11.042},
   journal={Physica A: Statistical Mechanics and its Applications},
   publisher={Elsevier BV},
   author={Harvey, M. and Hendricks, D. and Gebbie, T. and Wilcox, D.},
   year={2017},
   month=Apr, pages={416–426} }

@misc{List_corp,
  author       = {{Listcorp}},
  title        = {Johannesburg Stock Exchange (JSE) Listed Companies},
  year         = {2026},
  howpublished = {\url{https://www.listcorp.com/jse}},
  note         = {Accessed: 17 March 2026}
}

@book{numericalrecipes,
  author    = {William H. Press and Saul A. Teukolsky and William T. Vetterling and Brian P. Flannery},
  title     = {Numerical Recipes: The Art of Scientific Computing},
  edition   = {3},
  year      = {2007},
  publisher = {Cambridge University Press},
  address   = {Cambridge}
}

@misc{goliath_github_repo,
  author       = {Goliath, Ezra and Gebbie, Tim},
  title        = {Metaorder modelling and identification (GitHub repository)},
  year         = {2026},
  howpublished = {\url{https://github.com/EzraGoliath/Metaorder-modelling-and-identification-Msc-thesis-}},
  note         = {Accessed: 2026-02-17; repository has no archived release cited}
}



\section{Appendix}\label{sec:appendix}

\subsection{Legacy approach}\label{app:legacy approach}

The Version 1 implementation used the top 100 stocks (by market capitalisation) and closely followed the methodology of \citeauthor{Maitrieretal2025}. Parameter values for configurations were chosen to closely match the ones they had used and were not calibrated to the JSE market data we used. As a result, Version 1 served primarily as a proof of concept. 

We identified three key issues with our Version 1 implementation \footnote{We thank J-P Bouchaud for thoughtful comments and suggestions with regards to our initial v1 working paper.}. The first issue was that the execution profiles for GFI and GRT were too flat. Second, the fitted impact decay function for GRT had the correct shape, but the measured values were extremely noisy. Thirdly, we were able to recover the $\gamma \approx \alpha - 1$ relation, but it was crude and had a lot of noise. Our hypothesis for all these issues was that they were either due to a misspecification of the trader participation distribution, a misspecification of the number of traders $N$, a misspecification of the power-law exponent $\delta$, or a combination of these possible issues.

\subsubsection{Issues with the execution profiles}

When the metaorder impact (or any scaled version of the impact) is plotted as a function of the proportion of the metaorder that has been executed, $\phi$, we expect to see a concave relationship. This stylised fact of metaorder impact has been empirically shown \citep{2014marketimpactslifecycle, brokmann2014slowdecayimpactequity, zarinelli2014}.

\begin{figure*}[h!]
    \centering
    \begin{subfigure}[h]{0.45\linewidth}
        \includegraphics[width=\textwidth]{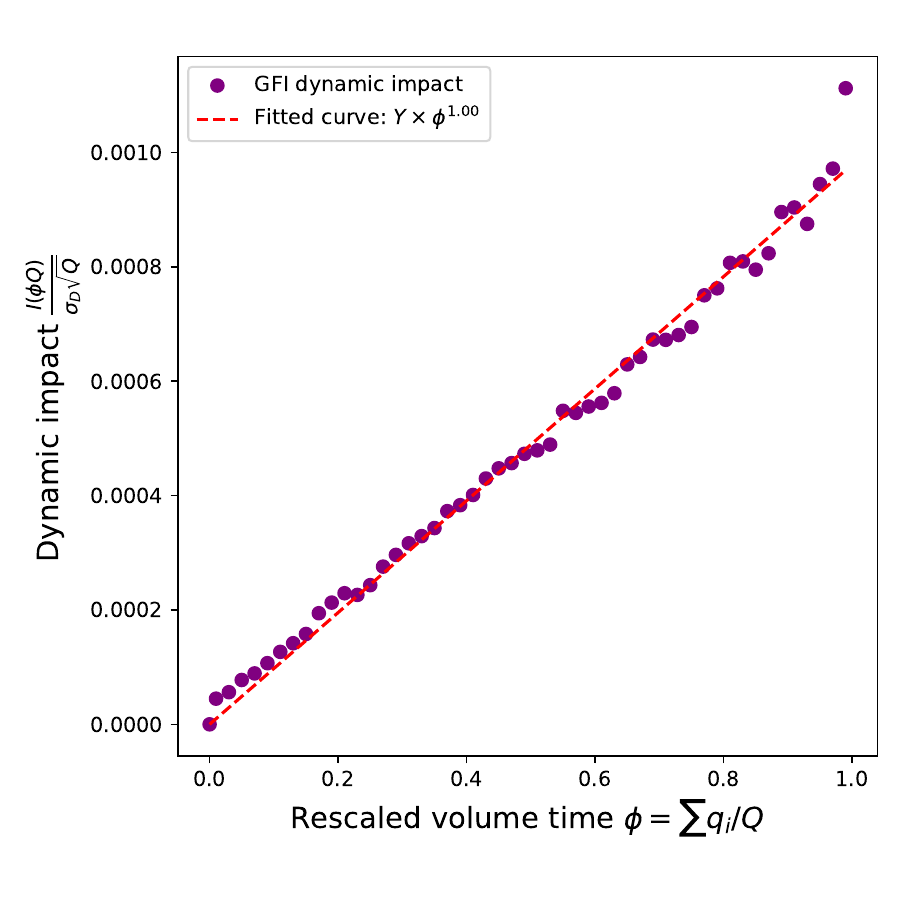}
        \caption{GFI}
        \label{subfig:GFI concave profile (legacy)}
    \end{subfigure}
    \begin{subfigure}[h]{0.45\linewidth}
        \includegraphics[width=\textwidth]{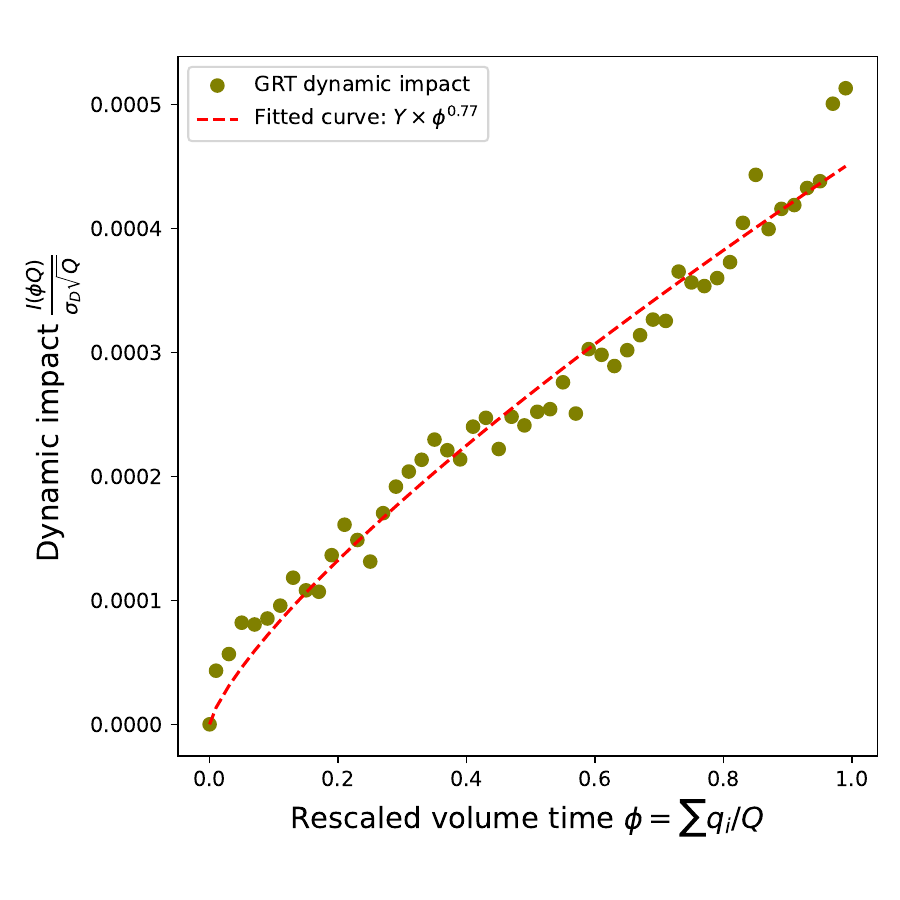}
        \caption{GRT}
        \label{subfig:GRT concave profile (legacy)}
    \end{subfigure}
    \caption{Concave profile of metaorder impact for GFI (a) and GRT (b) when using the Version 1 implementation. Metaorders were generated by specifying 20 traders and a power-law trader participation distribution with $\delta = 2$. Only metaorders with 10 or more child orders were used for this analysis. The fitted curves are shown in red. See Table \ref{tab:concave profiles (legacy)} for the fitted profiles for GFI and GRT respectively.}
    \label{fig:concave profiles (legacy)}
\end{figure*} 

\begin{table}[h!]
\centering
\begin{tabular}{lccc}
\toprule
Description & Ticker & \makecell{prefactor \\ $[\times10^{-4}]$}  & exponent \\
\midrule
Gold Fields Ltd. & GFI & 9.79 $\pm$  0.20 & 1.001 $\pm$ 0.044\\
Growthpoint Ltd. & GRT &  4.54 $\pm$ 0.14 & 0.766 $\pm$ 0.057\\ 
\bottomrule
\end{tabular}
\caption{Fitted values for metaorder execution profiles of GFI and GRT for the prefactor, and the exponent of phi. These are the fits for Figure \ref{subfig:GFI concave profile (legacy)} for GFI and Figure \ref{subfig:GRT concave profile (legacy)} for GRT.}
\label{tab:concave profiles (legacy)}
\end{table}

From Figure \ref{fig:concave profiles (legacy)} and Table \ref{tab:concave profiles (legacy)}, it is clear that the execution profile for GFI is flat. The exponent of phi is 1 and indicates that the impact grows linearly in $\phi$. This result contradicts the results in \citep{Maitrieretal2025} which used synthetic metaorders and the results in \citep{2014marketimpactslifecycle, brokmann2014slowdecayimpactequity, zarinelli2014} which used real meta\-orders.

\subsubsection{Issues with the decay profile}

\begin{figure}[h!]
    \centering
    \includegraphics[width=0.9\linewidth]{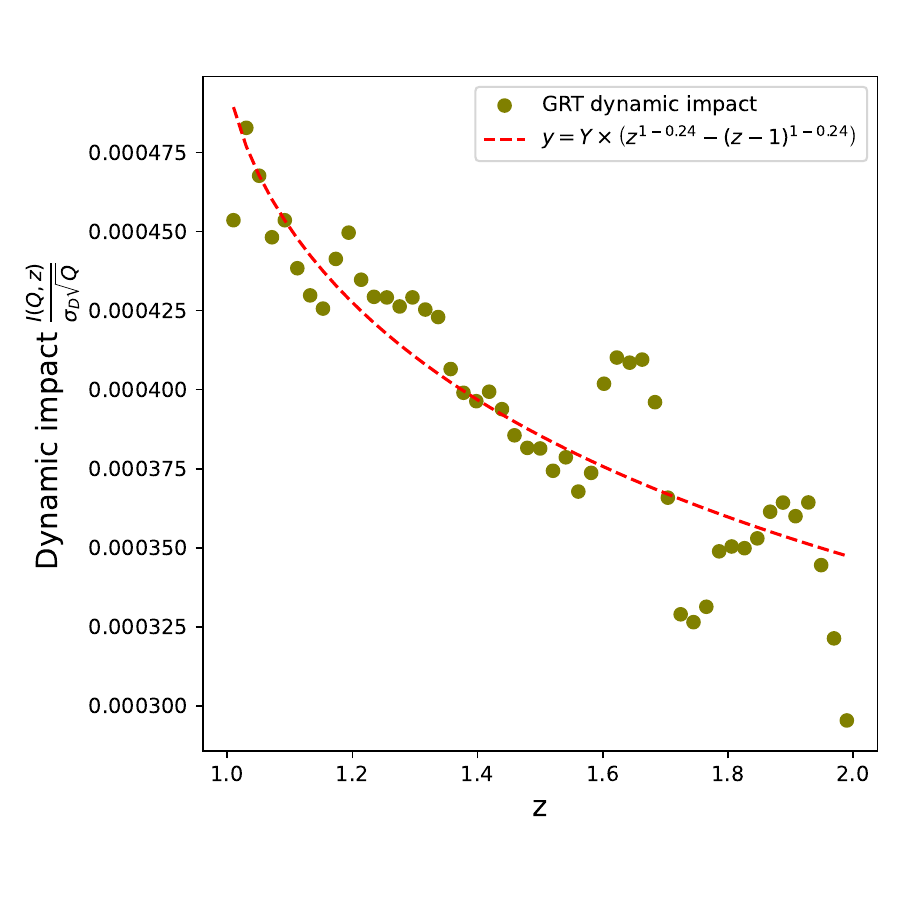}
    \caption{Post execution decay of impact for GRT. Dynamic impact is plotted as a function of $z$, the rescaled time. Metaorders were generated by specifying 20 traders and a power-law trader participation distribution with $\delta = 2$. The fitted line is shown in red.}
    \label{fig:impact decay (legacy)}
\end{figure}

\citeauthor{Maitrieretal2025} were able to use their synthetic metaorders to recover a concave decay profile. Using their methodology, we were able to recover $\beta = 0.241 \pm 0.033$ (Table \ref{tab:impact decay (legacy)}) which is similar to the $\beta = 019$ found in \cite{Maitrieretal2025} which used synthetic metaorders and $\beta = 0.2$ found in \citep{brokmann2014slowdecayimpactequity} which used real metaorders. However, from Figure \ref{fig:impact decay (legacy)} it is clear that while the shape is correct, there is a lot of noise in the measured decay. 

\begin{table}
\centering
\begin{tabular}{lccc}
\toprule
Description & Ticker & \makecell{prefactor \\ $[\times10^{-4}]$} & $\beta$ \\
\midrule
Growthpoint Ltd.& GRT & 5.01 $\pm$ 0.194 & 0.241 $\pm$ 0.033 \\
\bottomrule
\end{tabular}
\caption{Fitted values for metaorder post execution impact decay for GRT. }
\label{tab:impact decay (legacy)}
\end{table}

\subsubsection{Issues with the LMF relation}

\citeauthor{LilloMikeFarmer2005} proposed a microscopic explanation for the long-range correlation that can be observed from the market-order flow. They relate the power-law exponent of the metaorder run lengths $\alpha$, to the exponent of power-law decay decay of the autocorrelation function of the trade signs $\gamma$. Specifically, they propose the relation $\gamma \approx \alpha - 1$. 
This relation has been quantitatively tested using account-level order-splitting data from the Tokyo Stock Exchange (TSE) \cite{SatoKanazawa2023LMFTest}. 

\begin{figure*}[h!]
    \centering
    \begin{subfigure}[h]{0.45\linewidth}
        \includegraphics[width=\textwidth]{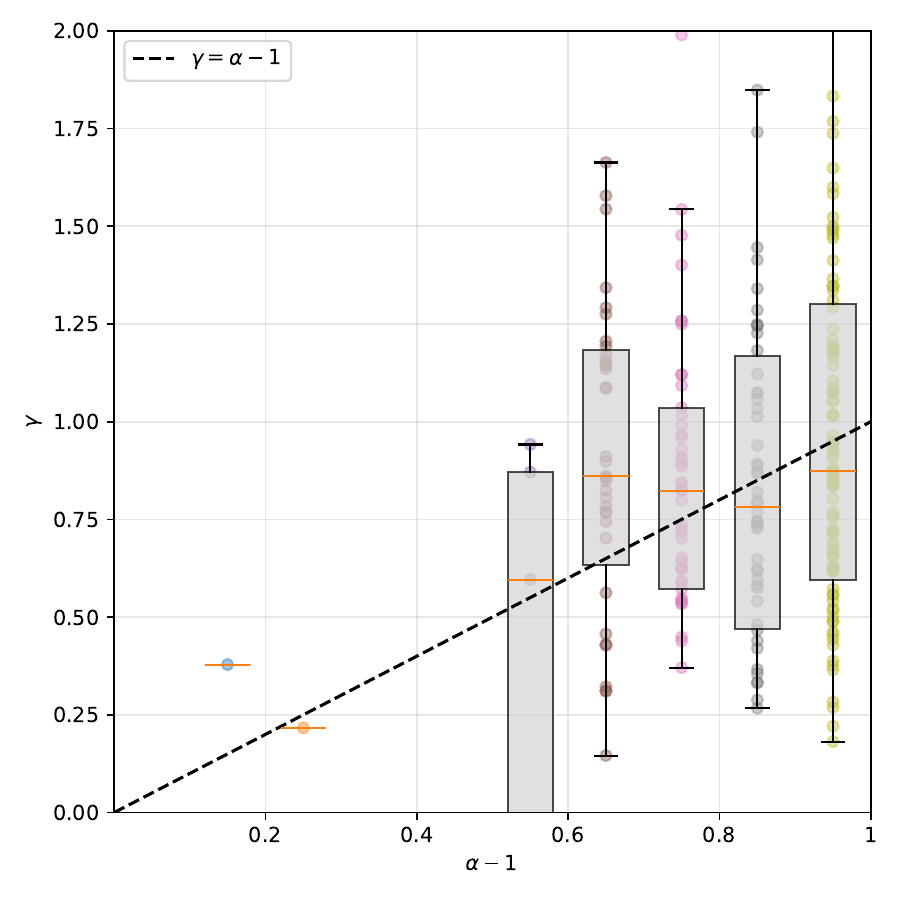}
        \caption{N = 50}
        \label{subfig:alpha gamma 50}
    \end{subfigure}
    \begin{subfigure}[h]{0.45\linewidth}
        \includegraphics[width=\textwidth]{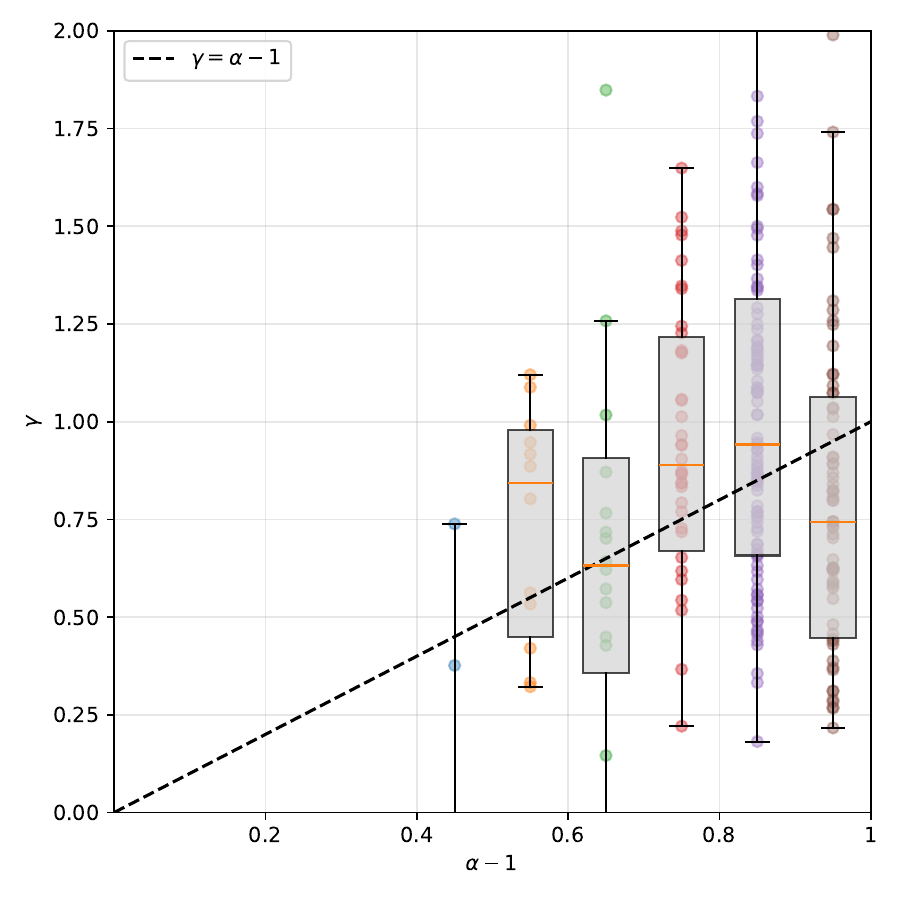}
        \caption{N = 150}
        \label{subfig:alpha gamma 150}
    \end{subfigure}
    \caption{Scattered box plots between $\alpha -1$ and $\gamma$ with the median indicated in orange. (a) shows the results when using a power-law trader distribution with $N=50$ and $\delta$ = 2 and (b) shows the result when using a power-law distribution with $N=150$ and $\delta = 2$. In both cases, the NLLS method was used to estimate $\gamma$.}
    \label{fig:alpha gamma (legacy)}
\end{figure*}

From Figure \ref{fig:alpha gamma (legacy)}, we see that the relation is recovered. The relation is however quite crude. The bars are large and the whiskers are long. The results we obtained did not match the results obtained in \cite{SatoKanazawa2023LMFTest} and indicate that there could be a misspecification in $N$ or $\delta$.

\subsection{Stocks used in the analysis} \label{app:data-metadata}

\begin{table}[h!]
\begin{tabular}{|c|c|c|c|c|c|}
\hline
4SI & ABG & ACL & ACS & ACT & ADH \\ \hline
ADR & ADW & AEG & AEL & AFE & AFH \\ \hline
AFT & AGL & AME & ANG & ANH & AON \\ \hline
AOO & APF & APH & APN & ARI & ARL \\ \hline
ART & ATT & AVI & BAT & BCF & BEL \\ \hline
BHG & BID & BIK & BLU & BOX & BRN \\ \hline
BRT & BTI & BVT & BWN & BYI & CAA \\ \hline
CAC & CAT & CCO & CFR & CGR & CHP \\ \hline
CKS & CLH & CLI & CLS & CMH & CML \\ \hline
CPI & CSB & CTA & CVW & DCP & DIB \\ \hline
DLT & DNB & DRD & DSY & DTC & EEL \\ \hline
EMH & EMI & EMN & ENX & EOH & EPE \\ \hline
EPS & EQU & EUZ & EXP & EXX & FBR \\ \hline
FFB & FGL & FSR & GFI & GLI & GLN \\ \hline
GML & GND & GPL & GRT & GTC & HAR \\ \hline
HCI & HDC & HET & HIL & HLM & HMN \\ \hline
HUG & HYP & IMP & INL & INP & IPF \\ \hline
ISA & ISB & ISO & ITE & IVT & JBL \\ \hline
JSE & KAL & KAP & KBO & KIO & KP2 \\ \hline
KRO & KST & LAB & LBR & LEW & LHC \\ \hline
LTE & MCZ & MDI & MMP & MNP & MPT \\ \hline
MRF & MRP & MSP & MST & MTA & MTH \\ \hline
MTM & MTN & N91 & NCS & NED & NPH \\ \hline
NPK & NPN & NRL & NRP & NTC & NVS \\ \hline
NWL & NY1 & OAO & OAS & OCE & OCT \\ \hline
OMN & OMU & OPA & ORN & OUT & PAN \\ \hline
PBG & PIK & PMR & PMV & PPC & PPE \\ \hline
PPH & PPR & PRX & QFH & QLT & RBO \\ \hline
RBX & RCL & RDF & REM & RES & RFG \\ \hline
RHB & RLO & RMH & RNG & RNI & RTN \\ \hline
RTO & S32 & SAC & SAP & SBK & SBP \\ \hline
SCD & SDL & SDO & SEA & SEB & SEP \\ \hline
SHG & SHP & SLM & SNT & SNV & SOH \\ \hline
SOL & SOLBE1 & SPG & SPP & SRE & SRI \\ \hline
SSK & SSS & SSW & SUI & SUR & SYG \\ \hline
TBS & TCP & TDH & TEX & TFG & TGA \\ \hline
THA & TKG & TLM & TMT & TPC & TRL \\ \hline
TRU & TSG & TTO & UPL & VAL & VIS \\ \hline
VKE & VOD & VUN & WBC & WBO & WEZ \\ \hline
WHL & YRK & YYLBEE & ZED & ZZD &  \\ \hline
\end{tabular}
\caption{The JSE tickers of the 239 stocks used in the analyses for the project. These stocks were selected from the top 250 JSE-listed companies by market capitalisation as of 13 March 2026 \citep{List_corp}. Of these 250 companies, data for 239 were available through BMLL and were therefore included in the analyses \citep{bmllpython2026}. Additional information, including company names, descriptions, and relevant links are available in the project repository \citep{goliath_github_repo}. See Table \ref{tab:processed data sample} for an example of the panel of processed data (after the data engineering) that is ready for the datascience.}
\label{tab:239 tickers}
\end{table}

The 239 stocks used in the analyses are shown in Table \ref{tab:239 tickers}. The table displays the tickers as presented by the JSE. These stocks were chosen as they represent the Top 250 stocks on the JSE (sorted by market cap) according to \citep{List_corp} as of 13 March 2026. Only 239 of the 250 were available on BMLL \citep{bmllpython2026}. 

\subsection{Sample of the processed data}\label{app:data-sample processed data}

Table \ref{tab:processed data sample} reports a snippet of the raw data after some basic processing. The data from BMLL \citep{bmllpython2026} was cleaned and had some basic processing done to it. The only processing done by us was the addition of the Mid-price after (delayed), Mid-price after (1ms)(delayed), Daily Volume, Intraday volatility, 20 AD volume and 20 AD volatility. {\it Mid-price after (immediate)} is the mid-price immediately after the trade (only this value was used in the analyses for calculating impact). {\it Daily Volume} is the total volume executed in the day. {\it Intraday volatility} is the daily volatility, calculated as in Algorithm \ref{alg:synthetic_metaorders}. {\it 20 AD volume} and {\it 20 AD volatility} are the 20 day average volume and volatilities respectively. \footnote{Columns that contain mid-prices followed by {\it (1ms)} indicate that the mid-price was measured 1 microsecond before/after execution began/ended. Columns without the {\it (1ms)} indicate that it is the last/first mid-price update in the order book before/after execution starts/ends. In most cases these two values are the same, but they differ in cases where there is no order book update within 1 microsecond of the execution starting/ending. Columns that contain the mid-prices after followed by {\it (delayed)} indicates that it is the mid-price after execution of the final child order of the metaorder and just before execution of the next trade. This value may be of use for investigating if the impact is transient or permanent but we have not used it in our analysis thus far} 

\subsection{Sample of data after applying algorithms}\label{app:data-sample metaorder}

Table \ref{tab:processed metaorder sample} shows a snippet of the processed data after applying algorithms \ref{alg:mapping} and \ref{alg:synthetic_metaorders}. We generated the synthetic metaorders by specifying 10 traders per month and a homogeneous trader participation distribution. The table shows the first 3 metaorders for Anglo American plc (Ticker AGL). 

The table reports multiple different versions of the impact of the metaorder and the differences between them amount to differences in timing. For our analysis $\it {impact}$ was used. This was calculated as the log difference in mid-price immediately after execution ends and just before execution begins.

\begin{sidewaystable}[t]
\centering
\small

\textbf{Panel A}

\begin{tabular}{rllrllrrrrr}
\toprule
Row & MIC & Ticker & ListingId & Date & DateTime & ExchangeSequenceNo & Trade Sign & Price & Volume &
\makecell{Mid-price\\before (1ms)} \\
\midrule
1 & XJSE & AGL & 418405540 & 2023-01-03 & 2023-01-03 09:14:19.712574 & 446408 & 1 & 66584.00 & 150.00 & 66553.00 \\
2 & XJSE & AGL & 418405540 & 2023-01-03 & 2023-01-03 09:14:20.108198 & 446486 & -1 & 66524.00 & 38.00 & 66554.00 \\
3 & XJSE & AGL & 418405540 & 2023-01-03 & 2023-01-03 09:14:25.318408 & 447996 & 1 & 66584.00 & 150.00 & 66554.00 \\
\bottomrule
\end{tabular}

\vspace{1cm}

\textbf{Panel B (continued)}

\begin{tabular}{rrrrrrrrrr}
\toprule
Row &
\makecell{Mid-price\\after (1ms)} &
\makecell{Mid-price\\after (1ms)(delayed)} &
\makecell{Mid-price\\before} &
\makecell{Mid-price\\after} &
\makecell{Mid-price\\after (delayed)} &
Intraday Volatility &
Daily Volume &
20 AD volume &
20 AD volatility \\
\midrule
1 & 66554.00 & 66554.00 & 66553.00 & 66553.50 & 66554.00 & 0.03 & 449528.00 & 449528.00 & 0.03 \\
2 & 66554.00 & 66554.00 & 66554.00 & 66554.00 & 66554.00 & 0.03 & 449528.00 & 449528.00 & 0.03 \\
3 & 66597.00 & 66554.00 & 66554.00 & 66554.00 & 66554.00 & 0.03 & 449528.00 & 449528.00 & 0.03 \\

\bottomrule
\end{tabular}

\caption{Sample of processed data for Anglo American plc (Ticker AGL)}
\label{tab:processed data sample}
\end{sidewaystable}
  
\begin{sidewaystable}[t]
\centering
\small
\textbf{Panel A}

\begin{tabular}{r lrrllrrrrll}
\toprule
Row &
Date &
trader id &
metaorder id &
Ticker &
MIC &
\makecell{Daily\\Volume} &
\makecell{Intraday\\Volatility} &
\makecell{20 AD\\volume} &
\makecell{20 AD\\volatility} &
\makecell{Start\\time} &
\makecell{End\\time} \\
\midrule
1 & 2023-01-03 & 7 & 0 & AGL & XJSE & 449528.00 & 0.03 & 449528.00 & 0.03 & 2023-01-03 09:14:19.712574 & 2023-01-03 09:18:35.071031 \\
2 & 2023-01-03 & 2 & 0 & AGL & XJSE & 449528.00 & 0.03 & 449528.00 & 0.03 & 2023-01-03 09:16:10.079238 & 2023-01-03 09:24:17.372891 \\
3 & 2023-01-03 & 9 & 2 & AGL & XJSE & 449528.00 & 0.03 & 449528.00 & 0.03 & 2023-01-03 09:17:00.760830 & 2023-01-03 09:20:41.055928 \\
\bottomrule
\end{tabular}

\vspace{0.8cm}

\textbf{Panel B (continued)}

\begin{tabular}{r rrr rrrrrr}
\toprule
Row &
\makecell{volume\\traded} &
\makecell{number child\\orders} &
\makecell{trade\\sign} &
\makecell{Mid-price\\before} &
\makecell{Mid-price\\after} &
\makecell{Mid-price after\\(delayed)} &
\makecell{Mid-price before\\(1ms)} &
\makecell{Mid-price after\\(1ms)} &
\makecell{Mid-price after\\(1ms)(delayed)} \\
\midrule
1 & 309.00 & 3 & 1 & 66553.00 & 66709.00 & 66704.50 & 66553.00 & 66707.00 & 66704.50 \\
2 & 1252.00 & 5 & -1 & 66730.00 & 66552.50 & 66552.50 & 66730.00 & 66509.00 & 66475.50 \\
3 & 791.00 & 8 & -1 & 66754.00 & 66463.00 & 66465.00 & 66754.00 & 66461.50 & 66465.00 \\
\bottomrule
\end{tabular}

\vspace{0.8cm}

\textbf{Panel C (continued)}

\begin{tabular}{r rrrrr}
\toprule
Row &
impact &
\makecell{impact\\(delayed)} &
\makecell{impact\\(1ms)} &
\makecell{impact\\(1ms)(delayed)} \\
\midrule
1 & 0.00 & 0.00 & 0.00 & 0.00 \\
2 & 0.00 & 0.00 & 0.00 & 0.00 \\
3 & 0.00 & 0.00 & 0.00 & 0.00 \\
\bottomrule
\end{tabular}

\caption{Sample of processed data for Anglo American plc (Ticker AGL) after applying algorithms 1 and 2}
\label{tab:processed metaorder sample}
\end{sidewaystable}

\end{document}